\documentclass[10pt,aps,showpacs,nofootinbib,prd,aps,epsf,floats,
               amsmath,amssymb,amsfonts,axodraw]{revtex4}
\usepackage{amsmath, amssymb}
\newcommand{\mathsym}[1]{{}}

\usepackage{graphicx}
\usepackage{amsmath}
\usepackage{amssymb}
\usepackage{bm}
\newcommand{\rem}[1]{}
\newsavebox{\PSLASH}
 \sbox{\PSLASH}{$p$\hspace{-1.8mm}/}
 
\renewcommand{\theequation}{\thesection.\arabic{equation}}
\newcounter{saveeqn}
\newcommand{\add}{\addtocounter{equation}{1}}
\newcommand{\alpheqn}{\setcounter{saveeqn}{\value{equation}}%
\setcounter{equation}{0}%
\renewcommand{\theequation}{\mbox{\thesection.\arabic{saveeqn}{\alph{equation}}}}}
\newcommand{\reseteqn}{\setcounter{equation}{\value{saveeqn}}%
\renewcommand{\theequation}{\thesection.\arabic{equation}}}

 \newsavebox{\notrightarrow}
 \sbox{\notrightarrow}{$\to$\hspace{-4mm}/}
 
 \newsavebox{\PARTIALSLASH}
 \sbox{\PARTIALSLASH}{$\partial$\hspace{-1.6mm}/}
 
 \newsavebox{\ASLASH}
 \sbox{\ASLASH}{$A$\hspace{-2.1mm}/}
 
 \newsavebox{\KSLASH}
 \sbox{\KSLASH}{$k$\hspace{-1.8mm}/}
 
 \newsavebox{\LSLASH}
 \sbox{\LSLASH}{$\ell$\hspace{-1.8mm}/}
 
 \newsavebox{\QSLASH}
 \sbox{\QSLASH}{$q$\hspace{-1.8mm}/}
 
 \newsavebox{\DSLASH}
 \sbox{\DSLASH}{$D$\hspace{-2.2mm}/}

 \newcommand{\blue}{\IfColor{\textCadetBlue}{}}
\newcommand{\black}{\IfColor{\textBlack}{}}
\newcommand{\red}{\IfColor{\textRed}{}}
\newcommand{\green}{\IfColor{\textOliveGreen}{}}
\newcommand{\lila}{\IfColor{\textRedViolet}{}}

\begin{document}
\title{The effect of magnetization and electric polarization\\ on the anomalous transport coefficients of a chiral fluid}
\author{N. Sadooghi}\email{sadooghi@physics.sharif.ir (Corresponding-Author)}
\author{S. M. A. Tabatabaee}\email{tabatabaeemehr_sma@physics.sharif.edu}
\affiliation{Department of Physics, Sharif University of Technology,
P.O. Box 11155-9161, Tehran-Iran}
\begin{abstract}
The effects of finite magnetization and electric polarization on dissipative and non-dissipative (anomalous) transport coefficients of a chiral fluid are studied.  First, using the second law of thermodynamics as well as Onsager's time reversal symmetry principle, the complete set of dissipative transport coefficients of this medium is derived. It is shown that the properties of the resulting shear and bulk viscosities are mainly affected by the anisotropy induced by external electric and magnetic fields.
Then, using the fact that the anomaly induced currents do not contribute to entropy production, the corresponding algebro-differential equations to non-dissipative anomalous transport coefficients are derived in a certain derivative expansion.
The solutions of these equations show that, within this approximation, anomalous transport coefficients are, in particular, given in terms of the electric susceptibility of the medium.
\end{abstract}
\maketitle
\section{Introduction}\label{Introduction}\label{sec1}
\par\noindent
Symmetry principles play an important role in physics. Although the laws of nature are often dictated by these principles, as it turns out, many physical phenomena are due to various mechanisms of symmetry breaking. The latter includes explicit, spontaneous and anomalous symmetry breaking mechanisms. Among them, anomalous symmetry breaking is a pure quantum mechanical phenomenon, which has no counterpart in classical (non-relativistic) physics. If a global symmetry is anomalous, it implies that the symmetry of the classical Lagrangian of the theory is not obeyed in the quantum theory. In other words, the Noether current associated with that anomalous global symmetry is no longer conserved on the quantum level. This anomaly can be determined, e.g. by putting the theory in a background electromagnetic field, which couples to this anomalous current, and probes in this way the properties of the medium.
\par
Recently, there have been many attempts to observe the effects of quantum chiral anomalies in experiments. In particular, heavy ion collisions (HICs) are of enormous interest, mainly because of the possible creation of very strong magnetic fields at an early stage of these collisions \cite{kharzeev2007, skokov2009}. The new state of matter, the plasma of quarks and gluons, which is produced at the same stage, is believed to include asymptotically free chiral fermions. The interplay between quantum anomalies with external magnetic fields results in a variety of novel non-dissipative anomalous transport phenomena in such a system (for a recent review see \cite{kharzeev2013, liao2014, liao2015} and the references therein). These phenomena can principally be observed in HIC experiments \cite{skokov2016, hattori2016}. \par
Some of these anomaly induced phenomena consist of chiral magnetic effect (CME) \cite{kharzeev2007, kharzeev2008}, chiral separation effect (CSE) \cite{son2004, metlitski2005}, chiral vortical effect (CVE) \cite{erdmenger2008, banerjee2008, torabian2009, son2009}, or chiral vortical separation effect (CVSE) \cite{liao2015}.
Whereas CME refers to the generation of an electric current induced by the chirality imbalance of the medium in the presence of an external magnetic field, CSE is characterized by an axial vector current which is generated along the external magnetic field. This is quite similar to the Ohm law, where an electric current is generated along an external electric field. In contrast to the Ohm's coefficient, however, the transport coefficients corresponding to CME and CSE currents are non-dissipative, and entirely dictated by chiral anomaly.
In analogy to the external magnetic field, a global rotation of the medium, quantified by its vorticity, leads, in a medium with non-vanishing electric and axial chemical potential, to non-dissipative vector and axial vector currents, which are proportional to the vorticity of the medium. The corresponding proportionality factors are referred to as CVE and CVSE  coefficients.
\par
In \cite{son2009, sadofyev2010}, these anomalous transport coefficients are determined within a relativistic hydrodynamical approach in the presence of an external magnetic field. Using the second law of thermodynamics and the fact that the anomaly induced currents do not contribute to entropy production \cite{kharzeev2011}, certain algebro-differential equations are derived, whose solutions yield the anomalous transport coefficients corresponding to the aforementioned effects. The main purpose of this paper is to extend the method introduced originally in \cite{son2009, sadofyev2010} to a medium with finite magnetization and electric polarization. To the best of our knowledge, these in-medium modifications of anomalous transport coefficients, including the linear response of the medium to external electromagnetic fields, are not yet studied in the literature. They seem, however, to be important not only from a theoretical point of view, but also because the experimentally relevant quark-gluon plasma turns out to have finite magnetic \cite{bali2012} and electric susceptibilities \cite{fukushima2010}.
\par
The present paper is organized as follows: In Sec. \ref{sec2}, we introduce the ideal electro-magnetohydrodynamical (EMHD) framework with finite magnetization and electric polarization. We essentially follow the method previously used in \cite{rischke2010, rischke2011}. However, in contrast to these works, where the effect of the external electric field is neglected, we consider the case of non-vanishing magnetic \textit{and} electric fields, and derive the relevant thermodynamic relations in the presence of finite magnetization \textit{and} electric polarization. Moreover, an anomalous current will be considered, which includes anomalous transport coefficient as in \cite{son2009, sadofyev2010}. This brings our derivation in connection with quantum anomalies, where parallel magnetic \textit{and} electric fields turn out to play a major role. In Sec. \ref{sec3A}, we first derive the dissipative transport coefficients of the anomalous EMHD by using the second law of thermodynamics and the Onsager's time reversal symmetry principle. In particular, we show that the dissipative part of the electric current, as well as the viscous stress tensor include a large number of thermal and electric conductivities as well as shear and bulk viscosities. Their properties are mainly affected by the anisotropies induced by external electric and magnetic fields. Our results are therefore a completion of the results presented in \cite{rischke2010, rischke2011}, where the dissipative coefficients arisen from the electric field in the dissipative currents are absent. Moreover, our results include certain, previously discarded, dissipative coefficients which arise from the interplay between external electric and magnetic fields [see (\ref{EE17})].  In Sec. \ref{sec3B}, we eventually use the fact that the anomaly induced currents do not contribute to entropy production, and derive the corresponding algebro-differential equations to anomalous transport coefficients of this medium. We show that, within a certain second-order derivative expansion, these equations include, in particular, the electric susceptibility of the medium.
We then follow the method introduced in \cite{son2009}, and determine the anomalous transport coefficients by solving the above mentioned equations analytically. The effect of gravitational anomaly \cite{landsteiner2011-1,landsteiner2011-2} is not considered in the present work, neither in the case of vanishing nor in the case of non-vanishing susceptibilities. Section \ref{sec4} is devoted to a summary and a number of concluding remarks.
\section{Ideal Anomalous Electro-Magnetohydrodynamics}\label{sec2}
\setcounter{equation}{0}\par\noindent
Electro-magnetohydrodynamics addresses all phenomena related
to the interaction of electric and magnetic fields with an electrically
conducting magnetized fluid.
An ideal and locally equilibrated relativistic fluid is
characterized by its long-wavelength degrees of freedom, the
four-velocity, the temperature and the chemical potential fields,
$u^{\mu}(x)$, $T(x)$ and $\mu(x)$, respectively. The four-velocity
$u^{\mu}=\gamma(1,\mathbf{v})$, with $\gamma\equiv \sqrt{1-\mathbf{v}^{2}}$, is defined by the variation of the four-coordinate
$x^{\mu}$ with respect to the proper-time $\tau$, and satisfies
$u_{\mu}u^{\mu}=1$.\footnote{In the present work,
$g^{\mu\nu}=\mbox{diag}\left(1,-1,-1,-1\right)$.} In the absence of
external electromagnetic fields, the physical observables, the
entropy and baryonic currents $s^{\mu}_{(0)}$ and $n_{b(0)}^{\mu}$,
as well as the fluid energy-momentum tensor
${\cal{T}}^{\mu\nu}_{F(0)}$, are expressed in terms of
$u^{\mu}$ as\footnote{The subscript $(0)$ denotes the case of ideal
fluid.}
\begin{eqnarray}\label{NE1}
s_{(0)}^{\mu}\equiv su^{\mu},\qquad n_{b(0)}^{\mu}\equiv
n_{b}u^{\mu},\qquad
{\cal{T}}_{F(0)}^{\mu\nu}\equiv\epsilon
u^{\mu}u^{\nu}-p\Delta^{\mu\nu},
\end{eqnarray}
with $\Delta^{\mu\nu}\equiv g^{\mu\nu}-u^{\mu}u^{\nu}$, the
projector onto the direction perpendicular to $u^{\mu}$. In
(\ref{NE1}), $\epsilon$, $p$, $s$, $n_{b}$ are local energy density,
thermodynamic pressure, entropy and baryon number densities of the
ideal fluid, respectively. In an ideal and locally equilibrated
fluid with no electromagnetic fields and sources, the
quantities presented in (\ref{NE1}) are conserved
\begin{eqnarray}\label{NE2}
\partial_{\mu}s_{(0)}^{\mu}=0,~~~
\partial_{\mu}n_{b(0)}^{\mu}=0,~~~
\partial_{\mu}{\cal{T}}^{\mu\nu}_{F(0)}=0.
\end{eqnarray}
In the presence of electromagnetic fields, however, the total energy-momentum tensor of the ideal fluid is to be modified as \cite{rischke2010, rischke2011}
\begin{eqnarray}\label{NE3}
T_{(0)}^{\mu\nu}=T_{F(0)}^{\mu\nu}+T_{EM}^{\mu\nu},
\end{eqnarray}
where the fluid and electromagnetic energy-momentum tensors,
$T_{F(0)}^{\mu\nu}$ and $T_{EM}^{\mu\nu}$, read as
\begin{eqnarray}\label{NE4}
T_{F(0)}^{\mu\nu}&=&{\cal{T}}_{F(0)}^{\mu\nu}-\frac{1}{2}
\left(M^{\mu\lambda}F_{\lambda}^{\,\,\nu}+M^{\nu\lambda}F_{\lambda}^{~\mu}\right),\nonumber\\
T_{EM}^{\mu\nu}&=&-F^{\mu\lambda}F_{\hspace{0.2cm}\lambda}^{\nu}+\frac{1}{4}g^{\mu\nu}F^{\rho\sigma}F_{\rho\sigma}.
\end{eqnarray}
where ${\cal{T}}^{\mu\nu}_{F(0)}$, defined in (\ref{NE1}), is the energy-momentum tensor of an ideal un-magnetized and un-electropolarized fluid.\footnote{
It is important to notice that the subscript $(0)$ on $T_{F(0)}^{\mu\nu}$ in (\ref{NE4}) refers only to the zeroth-order hydrodynamical derivative expansion of the fluid part of the energy-momentum tensor, ${\cal{T}}^{\mu\nu}_{F(0)}=\epsilon u^{\mu}u^{\nu}-p\Delta^{\mu\nu}$ from (\ref{NE1}). Later we will add higher order terms in the derivative expansion to this part of $T^{\mu\nu}_{(0)}$ from (\ref{NE3}), which will contribute to dissipation. We will then determine dissipative coefficients up to first-order derivative expansion.}
The antisymmetric field strength and polarization tensors,
$F^{\mu\nu}$ and $M^{\mu\nu}$ in (\ref{NE4}), are expressed in terms of electric
and magnetic fields, $E$ and $B$, as
\begin{eqnarray}\label{NE5}
F^{\mu\nu}&\equiv& Ee^{\mu\nu}-Bb^{\mu\nu},
\nonumber\\
M^{\mu\nu}&\equiv& -Pe^{\mu\nu}-Mb^{\mu\nu},
\end{eqnarray}
with
\begin{eqnarray}\label{NE6}
e^{\mu\nu}\equiv e^{\mu}u^{\nu}-e^{\nu}u^{\mu},
\hspace{0.3cm}\mbox{and}\hspace{0.3cm}b^{\mu\nu}\equiv
\varepsilon^{\mu\nu\alpha\beta}b_{\alpha}u_{\beta},
\end{eqnarray}
as well as $e^{\mu}\equiv \frac{E^{\mu}}{E}$ and $b^{\mu}\equiv
\frac{B^{\mu}}{B}$. In a frame where the fluid is moving with the
velocity $u^{\mu}$, the four-vector of electric and magnetic fields
are given by $E^{\mu}\equiv F^{\mu\nu}u_{\nu}$ and
$B^{\mu}\equiv\frac{1}{2}
\varepsilon^{\mu\nu\alpha\beta}F_{\nu\alpha}u_{\beta}$. Here,
$\varepsilon^{\mu\nu\alpha\beta}$ is the totally antisymmetric
Levi-Civita tensor. In the rest frame of the fluid, with
$u^{\mu}=(1,\mathbf{0})$, we have therefore
$E^{\mu}=(0,{\mathbf{E}})$ and $B^{\mu}=(0,\mathbf{B})$. Here,
$E^{i}=F^{i0}$ and $B^{i}=-\varepsilon^{ijk}F_{jk}/2$, as in
non-relativistic electrodynamics. The strength of the electric and
magnetic fields, $E$ and $B$, are given by the normalization
relations $E^{\mu}E_{\mu}=-E^{2}$ and $B_{\mu}B^{\mu}=-B^{2}$, which
lead immediately to $e^{\mu}e_{\mu}=b^{\mu}b_{\mu}=-1$.
\par
The antisymmetric polarization tensor $M^{\mu\nu}$ in (\ref{NE5})
describes the response of the system to an applied field strength
$F^{\mu\nu}$, and leads through the relation $
F^{\mu\nu}-M^{\mu\nu}\equiv H^{\mu\nu}$ to the induced field
strength tensor $H^{\mu\nu}$, which in terms of $e^{\mu\nu}$ and
$b^{\mu\nu}$, is given by
\begin{eqnarray}\label{NE7}
H^{\mu\nu}\equiv De^{\mu\nu}-Hb^{\mu\nu}.
\end{eqnarray}
Here, in analogy to $E^{\mu}$ and $B^{\mu}$, we define the
four-vectors of induced electric and magnetic fields, $D^{\mu}\equiv
H^{\mu\nu}u_{\nu}$ as $H^{\mu}\equiv
\frac{1}{2}\varepsilon^{\mu\nu\alpha\beta}H_{\nu\alpha}u_{\beta}$.
They are in relation to the four-vectors of electric polarization
$P^{\mu}\equiv -M^{\mu\nu}u_{\nu}$ and
magnetization $M^{\mu}\equiv
\frac{1}{2}\varepsilon^{\mu\nu\alpha\beta}M_{\nu\alpha}u_{\beta}$.
In the rest frame of the fluid, $D^{\mu}=(0,\mathbf{D})$ and
$H^{\mu}=(0,\mathbf{H})$. Similarly, we have
$P^{\mu}=(0,\mathbf{P})$ and $M^{\mu}=(0,\mathbf{M})$. In
(\ref{NE5}) and (\ref{NE7}), $V\equiv |\mathbf{V}|$,
with $V=\{E,B,P,M,D,H\}$ is used. This fixes the normalization
relations $V^{\mu}V_{\mu}=-V^{2}$, with $V=\{E,B,P,M,D,H\}$.
\par
The purpose of this paper is to study the effect of electric
polarization $P$ and magnetization $M$ on dissipative and non-dissipative transport
coefficients of an electromagnetized chiral fluid affected by the quantum anomaly. To this purpose, we introduce the electric
and magnetic susceptibilities, $\chi_{e}$ and $\chi_{m}$, arising in
linearized relations between ${\mathbf{P}}$ and ${\mathbf{E}}$ as
well as ${\mathbf{M}}$ and ${\mathbf{B}}$,
${\mathbf{P}}=\chi_{e}{\mathbf{E}}$ and
$\mathbf{M}=\chi_{m}\mathbf{B}$. Moreover, to bring our derivation
in connection with quantum anomalies, we will assume,
\begin{eqnarray}\label{NE8}
e_{\mu}b^{\mu\nu}=0.
\end{eqnarray}
In the rest frame of the fluid, (\ref{NE8}) is equivalent to parallel electric and magnetic fields,
$\mathbf{e}\|\mathbf{b}$ with $\mathbf{e}\equiv \mathbf{E}/E$ and
$\mathbf{b}\equiv \mathbf{B}/B$ denoting the directions of external
electric and magnetic fields. Apart from (\ref{NE8}),
$e_{\nu}\partial_{\mu}b^{\mu\nu}=-b^{\mu\nu}\partial_{\mu}e_{\nu}=0$
and $\partial_{\mu}\chi_{e}=\partial_{\mu}\chi_{m}=0$ are assumed.
\par
In what follows, the in-medium Maxwell equations will be used to determine the
relevant thermodynamic relations for an ideal and locally
equilibrated fluid. A number of useful relations will be also
derived. We will keep our notations similar to what is presented in
\cite{rischke2010}, where, in contrast to our presentation, the
effect of electric field is neglected.
\par
Introducing, in analogy to $n_{b(0)}^{\mu}$ from (\ref{NE1}), the
four-vector of electric current, $n_{e(0)}^{\mu}\equiv n_{e}u^{\mu}$,
the inhomogeneous Maxwell equation reads
\begin{eqnarray}\label{NE9}
\partial_{\mu}H^{\mu\nu}=n_{e(0)}^{\nu},
\end{eqnarray}
where $H^{\mu\nu}$ is defined in (\ref{NE7}) and
$n_{e(0)}^{\mu}\equiv (n_{e},{\mathbf{n}}_{e})$. Here, $n_{e}$ is
the electric charge density and ${\mathbf{n}}_{e}$ is the
corresponding electric current. As expected, (\ref{NE9}) is consistent
with the conservation relation of the electric current,
$\partial_{\mu}n_{e(0)}^{\mu}=0$. Together with the
conservation relations $\partial_{\mu}s^{\mu}_{(0)}=0$ and
$\partial_{\mu}n^{\mu}_{b(0)}=0$ from (\ref{NE2}), an ideal fluid in
the presence of electromagnetic fields is particularly described by
the conservation of the full energy-momentum tensor
$T^{\mu\nu}_{(0)}$ from (\ref{NE3}),
\begin{eqnarray}\label{NE10}
\partial_{\mu}T^{\mu\nu}_{(0)}=0.
\end{eqnarray}
Using $T^{\mu\nu}_{EM}$ from (\ref{NE4}) and  the inhomogeneous
Maxwell equation (\ref{NE9}), we arrive at
\begin{eqnarray}\label{NE11}
\partial^{\mu}T_{\mu\nu}^{EM}=n_{\mbox{\tiny{tot}}}^{\mu}F_{\mu\nu},
\end{eqnarray}
with
\begin{eqnarray}\label{NE12}
n_{\mbox{\tiny{tot}}}^{\mu}\equiv
n_{e(0)}^{\mu}+\partial_{\rho}M^{\rho\mu},
\end{eqnarray}
with $M^{\mu\nu}=F^{\mu\nu}-H^{\mu\nu}$. To derive (\ref{NE11}), we
have used the homogeneous Maxwell equation,
$\varepsilon^{\mu\nu\alpha\beta}\partial_{\beta}F_{\nu\alpha}=0$.
Plugging $F^{\mu\nu}$ from (\ref{NE5}) into this equation, and using the standard
relation
$\varepsilon^{\mu\nu\alpha\beta}\varepsilon_{\alpha\beta\rho\sigma}=-2(\delta^{\mu}_{\
\rho}\delta^{\nu}_{\ \sigma}-\delta^{\mu}_{\ \sigma}\delta^{\nu}_{\
\rho})$, the homogeneous Maxwell equation is equivalently given by
\begin{eqnarray}\label{NE13}
\partial_{\beta}(\varepsilon^{\mu\nu\alpha\beta}E_{\nu}u_{\alpha})+\partial_{\beta}(B^{\mu}u^{\beta}
-B^{\beta}u^{\mu})=0.
\end{eqnarray}
We use this equation to derive two useful relations, which play
important roles in the rest of this work. First, contracting
(\ref{NE13}) with $u_{\mu}$, we obtain
\begin{eqnarray}\label{NE14}
\partial_{\mu}B^{\mu}=u\cdot {\cal{D}}B+2E\cdot \omega,
\end{eqnarray}
where ${\cal{D}}\equiv u^{\mu}\partial_{\mu}$ and $a\cdot b\equiv
a^{\mu}b_{\mu}$. The four-vector $\omega^{\mu}$, on the right-hand
side (r.h.s.) of (\ref{NE14}), is the vorticity of the fluid,
defined by
\begin{eqnarray}\label{NE15}
\omega^{\mu}\equiv
\frac{1}{2}\varepsilon^{\mu\nu\rho\sigma}u_{\nu}\partial_{\rho}u_{\sigma}.
\end{eqnarray}
Following the steps described in Appendix \ref{appA}, another useful relation
\begin{eqnarray}\label{NE15a}
\partial_{\mu}E^{\mu}=u\cdot {\cal{D}}E-\frac{2(1-\chi_{m})}{(1+\chi_{e})}B\cdot\omega+\frac{n_{e}}{1+\chi_{e}},
\end{eqnarray}
can be derived [see (\ref{appA13a})-(\ref{appA16a}) for the proof of (\ref{appA6a}) and set $\partial_{\mu}\chi_{e}=0$ to arrive at (\ref{NE15a})].
Contracting further (\ref{NE13}) with $b^{\mu}$, and using
(\ref{NE8}) as well as $\epsilon^{\mu\nu\alpha\beta}b^{\mu}\partial_{\beta}(E_{\nu}u_{\alpha})=0$, we arrive at
another useful relation
\begin{eqnarray}\label{NE16}
{\cal{D}}\ln B+\theta-u^{\nu}b^{\mu}\partial_{\mu}b_{\nu}=0,
\end{eqnarray}
where $\theta\equiv
\partial_{\mu}u^{\mu}$. The last useful relation reads
\begin{eqnarray}\label{NE17}
{\cal{D}}\ln E+\theta-u_{\nu}e^{\mu}\partial_{\mu}e^{\nu}=0.
\end{eqnarray}
To derive (\ref{NE17}), let us first consider the energy-momentum tensor
(\ref{NE3}). Plugging $F^{\mu\nu}$ and $M^{\mu\nu}$ from
(\ref{NE5}) into $T_{F(0)}^{\mu\nu}$ and $T^{\mu\nu}_{EM}$ from
(\ref{NE4}), we arrive after some algebraic manipulations at
\begin{eqnarray}\label{NE18}
T^{\mu\nu}_{F(0)}&=&\epsilon' u^{\mu}u^{\nu}-p_{\perp}\Xi_{B}^{\mu\nu}+p_{\|}b^{\mu}b^{\nu}-EPe^{\mu}e^{\nu},\nonumber\\
T^{\mu\nu}_{EM}&=&\frac{1}{2}B^{2}\left(u^{\mu}u^{\nu}-\Xi_{B}^{\mu\nu}-b^{\mu}b^{\nu}\right)+\frac{1}{2}E^{2}
\left(u^{\mu}u^{\nu}-\Xi_{E}^{\mu\nu}-e^{\mu}e^{\nu}\right),
\end{eqnarray}
where, $\epsilon'\equiv \epsilon+EP$, $p_{\perp}\equiv p-BM$,
$p_{\|}\equiv p$,
$\Xi_{B}^{\mu\nu}\equiv\Delta^{\mu\nu}+b^{\mu}b^{\nu}$, and
$\Xi_{E}^{\mu\nu}\equiv \Delta^{\mu\nu}+e^{\mu}e^{\nu}$. We further
consider
$\partial^{\mu}T_{\mu\nu}^{EM}=n_{\mbox{\tiny{tot}}}^{\mu}F_{\mu\nu}$
from (\ref{NE11}). Contracting this relation with $u^{\nu}$,
plugging $T^{\mu\nu}_{EM}$ from (\ref{NE18}) into the left hand side (l.h.s.) of the
resulting expression, and using eventually (\ref{NE8}) as well as
$b^{\mu\nu}\partial_{\mu}e_{\nu}=0$ and $\partial_{\mu}\chi_{e}=0$, which lead to
\begin{eqnarray*}
(\partial_{\mu}M^{\mu\nu})E_{\nu}=-\chi_{e}E^{2}\left({\cal{D}}\ln E+\theta-u_{\nu}e^{\mu}\partial_{\mu}e^{\nu}\right),
\end{eqnarray*}
we arrive at
\begin{eqnarray*}
(1+\chi_{e})\left({\cal{D}}\ln
E+\theta-u_{\nu}e^{\mu}\partial_{\mu}e^{\nu}\right)=0.
\end{eqnarray*}
The latter yields (\ref{NE17}) for $\chi_{e}\neq -1$.
\par
To check the consistency of the thermodynamic relations including
electric and magnetic fields, let us contract (\ref{NE10}) with
$u^{\nu}$. Using (\ref{NE17}) and
$\partial_{\mu}T^{\mu\nu}_{F(0)}=-\partial_{\mu}T^{\mu\nu}_{EM}$,
which arises from (\ref{NE10}) combined with the definition of
$T^{\mu\nu}_{(0)}$ from (\ref{NE4}), it is possible to show that
$u_{\nu}\partial_{\mu}T^{\mu\nu}_{F(0)}=0$. Plugging then
$T^{\mu\nu}_{F(0)}$ from (\ref{NE18}) into this relation, and using
$u^{\mu}b_{\mu}=u^{\mu}e_{\mu}=0$ and
$u_{\nu}\partial_{\mu}u^{\nu}=b_{\nu}\partial_{\mu}b^{\nu}=e_{\nu}\partial_{\mu}e^{\nu}=0$,
we arrive eventually at
\begin{eqnarray}\label{NE19}
{\cal{D}}\epsilon+(\epsilon+p)\theta+M{\cal{D}}B+E{\cal{D}}P=0.
\end{eqnarray}
The relevant thermodynamic relation for ${\cal{D}}\epsilon$ is then derived
by using
\begin{eqnarray}\label{NE20}
\epsilon+p=Ts+\mu_{b}n_{b}+\mu_{e}n_{e},
\end{eqnarray}
where $\mu_{b}$ and $\mu_{e}$ are the chemical potentials related
to the baryon and electric number densities, $n_{b}$ and $n_{e}$.
Plugging $\epsilon+p$ from (\ref{NE20}) into (\ref{NE19}), and using
the conservation laws of baryon number density,
$\partial_{\mu}n^{\mu}_{b(0)}=0$, electric number density
$\partial_{\mu}n^{\mu}_{e(0)}=0$ and the entropy density current
$\partial_{\mu}s^{\mu}_{(0)}=0$, we arrive after some algebraic
manipulations at
\begin{eqnarray}\label{NE21}
{\cal{D}}\epsilon=T{\cal{D}}s+\mu_{b}{\cal{D}}n_{b}+\mu_{e}{\cal{D}}n_{e}-M{\cal{D}}B-E{\cal{D}}P,
\end{eqnarray}
which is consistent with the standard thermodynamic relation
$d\epsilon=Tds+\mu_{b}dn_{b}+\mu_{e}dn_{e}-MdB-EdP$ \cite{schwablstatistic}. Combining at
this stage (\ref{NE20}) with (\ref{NE21}), the Gibbs-{\cal{D}}uhem relation
is given by
\begin{eqnarray}\label{NE22}
{\cal{D}}p=s{\cal{D}}T+n_{b}{\cal{D}}\mu_{b}+n_{e}{\cal{D}}\mu_{e}+M{\cal{D}}B+E{\cal{D}}P.
\end{eqnarray}
To obtain the standard Gibbs-{\cal{D}}uhem relation in the presence of electric
and magnetic fields, we define $\tilde{p}\equiv p-EP$. Using this
definition in (\ref{NE22}), we arrive at
\begin{eqnarray}\label{NE23}
{\cal{D}}\tilde{p}=s{\cal{D}}T+n_{b}{\cal{D}}\mu_{b}+n_{e}{\cal{D}}\mu_{e}+M{\cal{D}}B-P{\cal{D}}E.
\end{eqnarray}
Let us reiterate that the main aim of this paper is to study the effect
of quantum (axial) anomaly on the EMHD equations once
electric polarization $P$ and magnetization $M$ of the underlying
fluid are not neglected. To do this, we introduce at this stage a
$U(1)$ axial vector current,
\begin{eqnarray}\label{NE24}
n_{a(0)}^{\mu}\equiv n_{a}u^{\mu},
\end{eqnarray}
which satisfies the classical conservation law
\begin{eqnarray}\label{NE25}
\partial_{\mu}n_{a(0)}^{\mu}=0,
\end{eqnarray}
in the chiral limit.\footnote{We assume that the quark matter, which is
effectively described  by our ideal fluid, consists of massless
quarks.} Denoting the chemical potential associated with $n_{a}$ by
$\mu_{a}$, the thermodynamic relations (\ref{NE20}) and (\ref{NE21})
turn out to be
\begin{eqnarray}\label{NE26}
\epsilon+p=Ts+\mu_{b}n_{b}+\mu_{e}n_{e}+\mu_{a}n_{a},
\end{eqnarray}
and
\begin{eqnarray}\label{NE27}
{\cal{D}}\epsilon&=&T{\cal{D}}s+\sum\limits_{i=\{b,e,a\}}\mu_{i}{\cal{D}}n_{i}-M{\cal{D}}B-E{\cal{D}}P,
\end{eqnarray}
respectively. In the next section, we will consider the axial anomaly of the axial vector current,
\begin{eqnarray}\label{NE28}
\partial_{\mu}n_{a}^{\mu}=-\frac{e^{2}}{8\pi^{2}}F^{\mu\nu}\tilde{F}_{\mu\nu}=-CE\cdot B,
\end{eqnarray}
with $\tilde{F}_{\mu\nu}\equiv
\frac{1}{2}\varepsilon_{\mu\nu\rho\sigma}F^{\rho\sigma}$ and
$C\equiv\frac{e^{2}}{4\pi^{2}}$, and study its effect on non-dissipative transport coefficients of an electromagnetized relativistic fluid.\footnote{
The sign in front of $C$ in (\ref{NE28}) is fixed by assuming
parallel electric and magnetic fields aligned in the third
direction. This leads, according to our definitions, to $F_{12}=-B$
and $F^{03}=E$.}
\section{Dissipative and anomalous transport coefficients of a chiral fluid}\label{sec3}
\setcounter{equation}{0}
\par\noindent
In the first part of this section, Sec. \ref{sec3A}, we will derive the complete set of dissipative transport coefficients in the
presence of electric and magnetic fields. To do this, we will follow the formalism of dissipative fluid dynamics from
\cite{rischke2010}. In Sec. \ref{sec3B}, by taking into account the fact
that the anomaly induced current is non-dissipative \cite{kharzeev2011}, we will then derive the algebro-differential equations leading to anomalous transport
coefficients. Similar equations are derived originally in \cite{son2009, sadofyev2010}, where the in-medium effects are neglected. These equations will then be solved  in a medium with vanishing (Sec. \ref{sec3B1}) and non-vanishing  (Sec. \ref{sec3B2}) electric and magnetic susceptibilities. In what follows, we will first derive the general structure of $\partial_{\mu}s^{\mu}$, with $s^{\mu}$ the current of the entropy density of a dissipative anomalous fluid.
\par
Following the method presented in \cite{rischke2010, rischke2011}, we start by introducing the first-order dissipative and non-dissipative corrections to the conserved quantities of the ideal
fluid, $T^{\mu\nu}_{(0)},n_{b(0)}^{\mu},n_{e(0)}^{\mu}$ and
$s_{(0)}^{\mu}$,
\begin{eqnarray}\label{EE1}
T^{\mu\nu}&=&T^{\mu\nu}_{(0)}+\tau^{\mu\nu},\nonumber\\
n_{b}^{\mu}&=&n_{b}u^{\mu}+j_{b}^{\mu},\nonumber\\
n_{e}^{\mu}&=&n_{e}u^{\mu}+j_{e}^{\mu},\nonumber\\
s^{\mu}&=&su^{\mu}+j_{s}^{\mu}+
D_{\omega}\omega^{\mu}+D_{B}B^{\mu}+D_{E}E^{\mu}.
\end{eqnarray}
Here, the total energy-momentum tensor of the ideal fluid,
$T^{\mu\nu}_{(0)}$, is defined in (\ref{NE3}) and (\ref{NE4}), and
$n_{b},n_{e}$ and $s$ are the baryonic and electric number densities
as well as the entropy density of the ideal fluid, respectively. The coefficients
$D_{\omega},D_{B}$ and $D_{E}$ in $s^{\mu}$ are associated with the
anomaly. The coefficients $D_{\omega}$ and $D_{B}$ are originally
introduced in \cite{son2009}, where the entropy density current
$s^{\mu}$ is expanded only in terms of the vorticity $\omega_{\mu}$
from (\ref{NE15}) and the external magnetic field $B^{\mu}$. Here,
in the presence of magnetic \textit{and} electric fields, we have
also considered the effect of the external electric field $E^{\mu}$, and
introduced $D_{E}$ as its coefficient. Later, we will explicitly show that $D_{E}=0$. This is also expected from symmetry arguments. To
consider the quantum anomaly of the dissipative electromagnetized
fluid, we shall also replace the axial vector current
$n_{a(0)}^{\mu}$ from (\ref{NE24}) with
\begin{eqnarray}\label{EE2}
n_{a}^{\mu}=n_{a}u^{\mu}+j_{a}^{\mu}.
\end{eqnarray}
Whereas in (\ref{EE1}), $\tau^{\mu\nu}, j_{b}^{\mu}$ and
$j_{e}^{\mu}$ are dissipative currents,\footnote{Later,  we will see that $j^{\mu}_{e}$ consists of a dissipative and a non-dissipative part \cite{son2009, sadofyev2010}.} the additional
(non-dissipative) current $j_{a}^{\mu}$ in (\ref{EE2}) is introduced
in analogy to the considerations in \cite{son2009, sadofyev2010},
and will later be used to determine the algebro-differential
equations leading to anomalous transport coefficients.
\par
Let us notice that $j_{i}^{\mu},
i=b,e,a$ are orthogonal to $u^{\mu}$, \textit{i.e.}
$u_{\mu}j_{i}^{\mu}=0,$ for $i=b,e,a$. Moreover, $\tau^{\mu\nu}$ is a symmetric  rank-two tensor, satisfying the orthogonality condition $u_{\mu}\tau^{\mu\nu}=0$.
\par
At this stage, we will use
the conservation relations associated with the anomalous EMHD
\begin{eqnarray}\label{EE3}
\partial_{\mu}T^{\mu\nu}=0,&\qquad&
\partial_{\mu}n_{b}^{\mu}=0,\nonumber\\
\partial_{\mu}n_{e}^{\mu}=0,&\qquad&\partial_{\mu}n_{a}^{\mu}=-CE\cdot
B,
\end{eqnarray}
together with the second law of thermodynamics,
$T\partial_{\mu}s^{\mu}\geq 0$, in order to arrive at an
appropriate relation for $T\partial_{\mu}s^{\mu}$. To do this, we shall first
consider the conservation relation (\ref{NE11}) for the
electromagnetic energy-momentum tensor $T_{EM}^{\mu\nu}$ from
(\ref{NE18}). Replacing the ideal electric current $n_{e(0)}^{\mu}$
on the r.h.s. of (\ref{NE11}) by $n_{e}^{\mu}$
from (\ref{EE1}), and contracting the resulting expression with
$u_{\nu}$, we arrive at
\begin{eqnarray}\label{EE4}
{\cal{D}}\ln E+\theta-u_{\nu}e^{\mu}\partial_{\mu}e^{\nu}=\frac{E\cdot
j_{e}}{(1+\chi_{e})E^{2}}.
\end{eqnarray}
This relation replaces (\ref{NE17}) of the ideal electromagnetized fluid. To derive (\ref{EE4}), the relations (\ref{NE16}), together with
\begin{eqnarray}\label{EE5}
u_{\nu}\partial_{\mu}T^{\mu\nu}_{EM}&=&E^{2}\left({\cal{D}}\ln E+\theta-u_{\nu}e^{\mu}\partial_{\mu}e^{\nu}\right),\nonumber\\
(\partial_{\mu}M^{\mu\nu})E_{\nu}&=&-\chi_{e}E^{2}\left({\cal{D}}\ln E+\theta-u_{\nu}e^{\mu}\partial_{\mu}e^{\nu}\right),
\end{eqnarray}
are used. Then, using the conservation of the total energy-momentum tensor
$\partial_{\mu}T^{\mu\nu}=0$ from (\ref{EE3}) with
$T^{\mu\nu}$ from
(\ref{EE1}), and following the same steps leading from
$u_{\nu}\partial_{\mu}T^{\mu\nu}_{(0)}=0$ to (\ref{NE19}), we obtain
\begin{eqnarray}\label{EE6}
{\cal{D}}\epsilon+(\epsilon+p)\theta
+M{\cal{D}}B+E{\cal{D}}P+E\cdot j_{e}+u_{\nu}\partial_{\mu}\tau^{\mu\nu}=0.
\end{eqnarray}
To arrive at (\ref{EE6}), the identity (\ref{EE4}) and
\begin{eqnarray}\label{EE7}
u_{\nu}\partial_{\mu}T^{\mu\nu}_{F(0)}&=&{\cal{D}}\epsilon+(\epsilon+p)\theta+M{\cal{D}}B+E{\cal{D}}P
+\frac{\chi_{e}}{1+\chi_{e}}E\cdot j_{e},\nonumber\\
u_{\nu}\partial_{\mu}T^{\mu\nu}_{EM}&=&\frac{E\cdot j_{e}}{1+\chi_{e}},
\end{eqnarray}
are used. Then, expressing $j^{\mu}_{s}$ as a linear combination of
$j_{b}^{\mu}, j_{e}^{\mu}$ and $j_{a}^{\mu}$, as in
\cite{rischke2010},
\begin{eqnarray}\label{EE8}
j_{s}^{\mu}=-\alpha_{b}j_{b}^{\mu}-\alpha_{e}j_{e}^{\mu}-\alpha_{a}j_{a}^{\mu},
\end{eqnarray}
and using (\ref{NE26}) to express $(\epsilon+p)$ in (\ref{EE6}) in
terms of all the other thermodynamic variables, we arrive at
\begin{eqnarray}\label{EE9}
{\cal{D}}\epsilon-\sum_{i\in\{b,e,a\}}\mu_{i}{\cal{D}}n_{i}-T{\cal{D}}s+M{\cal{D}}B+E{\cal{D}}P+\sum_{i\in\{b,e,a\}}\mu_{i}\partial_{\mu}(n_{i}u^{\mu})+
T\partial_{\mu}(su^{\mu})+E\cdot j_{e}-\tau^{\mu\nu}\partial_{\mu}u_{\nu}=0.\nonumber\\
\end{eqnarray}
The first four terms on
the l.h.s. of (\ref{EE9}) vanish by making use of (\ref{NE27}) from
ideal EMHD. Replacing $n_{i}u^{\mu}$ and $su^{\mu}$ on the
l.h.s. of (\ref{EE9}) by their definitions
$n_{i}u^{\mu}=n_{i}^{\mu}-j_{i}^{\mu}$ for $i=b,e,a$ and
$su^{\mu}=s^{\mu}-j_{s}^{\mu}-D_{\omega}\omega^{\mu}-D_{B}B^{\mu}-D_{E}E^{\mu}$
from (\ref{EE1}), and using the conservation relations for
$n_{i}^{\mu}, i=b,e,a$ from (\ref{EE3}), as well as the expansion of
$j_{s}^{\mu}$ in terms of the other dissipative currents from
(\ref{EE8}), we arrive after some straightforward algebraic
manipulations at
\begin{widetext}
\begin{eqnarray}\label{EE10}
T\partial_{\mu}s^{\mu}&=&\tau^{\mu\nu}w_{\mu\nu}+(\mu_{a}-T\alpha_{a})\partial_{\mu}j_{a}^{\mu}+(\mu_{b}-T\alpha_{b})\partial_{\mu}j_{b}^{\mu}+
(\mu_{e}-T\alpha_{e})\partial_{\mu}j_{e}^{\mu}-j_{a}^{\mu}T\nabla_{\mu}\alpha_{a}-j_{b}^{\mu}T\nabla_{\mu}\alpha_{b}\nonumber\\
&&-j_{e}^{\mu}(T\nabla_{\mu}\alpha_{e}+E_{\mu})
+\mu_{a}CE\cdot
B+T\partial_{\mu}(D_{\omega}\omega^{\mu}+D_{B}B^{\mu}+D_{E}E^{\mu}).
\end{eqnarray}
\end{widetext}
\par\noindent
Here, $\nabla_{\mu}\equiv\Delta_{\mu\nu}\partial^{\nu}$ and
$w^{\mu\nu}\equiv
\frac{1}{2}\left(\nabla^{\mu}u^{\nu}+\nabla^{\nu}u^{\mu}\right)$ are
introduced. To satisfy the positivity condition of
$T\partial_{\mu}s^{\mu}$, the expression on the r.h.s. of
(\ref{EE10}) is to be non-negative. This leads immediately to
\begin{eqnarray}\label{EE11}
\begin{array}{rclcccc}
\alpha_{i}&=&\frac{\mu_{i}}{T},&&\mbox{for}&&i=a,b,e,\\
\end{array}
\end{eqnarray}
and the general Ansatz
\begin{eqnarray}\label{EE12}
j_{b}^{\mu}&=&-\sigma_{b}^{\mu\nu}T\nabla_{\nu}\alpha_{b},\nonumber\\
j_{e}^{\mu}&=&-\sigma_{e}^{\mu\nu}(T\nabla_{\nu}\alpha_{e}+E_{\nu})+\kappa_{\omega}\omega^{\mu}+\kappa_{B}B^{\mu}
+\kappa_{E}E^{\mu},\nonumber\\
j_{a}^{\mu}&=&\xi_{\omega}\omega^{\mu}+\xi_{B}B^{\mu}+\xi_{E}E^{\mu},
\nonumber\\
\tau^{\mu\nu}&=&\eta^{\mu\nu\rho\sigma}w_{\rho\sigma},
\end{eqnarray}
where $\sigma_{b}^{\mu\nu},\sigma_{e}^{\mu\nu}$ and
$\eta^{\mu\nu\alpha\beta}$ include dissipative transport
coefficients, and $\kappa_{i}$ and $\xi_{i}, i=B,E,\omega$ are non-dissipative coefficients. The latter can be
expressed in terms of anomalous transport coefficients. Let us notice that the
dissipative transport coefficients $\sigma_{b}^{\mu\nu},
\sigma^{\mu\nu}_{e}$ and $\eta^{\mu\nu\rho\sigma}$ are orthogonal to $u^{\mu}$, and are symmetric under $\mu\leftrightarrow\nu$. In the next two sections, we will use the
Onsager's time-reversal principle to first determine $\sigma_{i}^{\mu\nu}, i=b,
e$ and $\tau^{\mu\nu}$ in terms of thermal conductivity as
well as longitudinal and transverse shear and bulk viscosities. This
will generalize the standard formulation of magnetohydrodynamics presented in
\cite{rischke2010} to the case of non-vanishing electric field. We will then consider the anomalous contributions to
$j_{e}^{\mu}$ and $j_{a}^{\mu}$ in (\ref{EE10}) proportional to
$\kappa_{i}$ and $\xi_{i}, i=B,E, \omega$, and by combining them  \cite{sadofyev2010}, we will arrive at the
algebro-differential equations leading to
$\kappa_{i}$ and $\xi_{i}, i=B,E, \omega$ as well as $D_{i},
i=B,E, \omega$ in a dense and hot quark matter in the presence of
constant $\mathbf{E}$ and $\mathbf{B}$ fields. This will generalize the results presented in \cite{son2009, sadofyev2010} to the case of a fluid with finite
magnetization ${M}$ and electric polarization  ${P}$.
\subsection{Dissipative currents of an anomalous chiral fluid}\label{sec3A}
\par\noindent
According to the Onsager's principle for transport coefficients
\cite{lifshitz}, the thermal conductivity,
$\sigma_{i}^{\mu\nu}, i=b,e$, corresponding to the diffusive fluxes
of the baryonic and electric number density $n_{b}$ and $n_{e}$
shall satisfy
\begin{eqnarray}\label{EE13}
\sigma_{i}^{\mu\nu}(E,B)=\sigma_{i}^{\nu\mu}(E,-B), \qquad i=b,e.
\end{eqnarray}
Moreover, $\sigma_{b/e}^{\mu\nu}$ have to satisfy the orthogonality condition $u_{\mu}\sigma_{i}^{\mu\nu}=0$ for $i=b,e$, and are to be symmetric under $\mu\leftrightarrow \nu$. The relation $e_{\mu}b^{\mu\nu}=0$ from (\ref{NE8}) is also to be taken into account.
\par
To build $\sigma^{\mu\nu}_{b/e}$, we expand it in terms of
independent irreducible rank-two tensors, which are built from
$u^{\mu},g^{\mu\nu},b^{\mu}$ and $e^{\mu}$ \cite{rischke2010}. The only relevant tensors that are compatible with the above mentioned conditions are thus given by
\begin{eqnarray}\label{EE14}
&&\Delta^{\mu\nu},~b^{\mu}b^{\nu},~e^{\mu}e^{\nu}.
\end{eqnarray}
Other rank-two tensors like $b^{\mu\nu},~e^{[\mu,}b^{\nu]}, \cdots$ are excluded because of the aforementioned conditions. Here, $A^{[\mu,}B^{\nu]}\equiv A^{\mu}B^{\nu}-A^{\nu}B^{\mu}$.
Introducing at this stage three
independent thermal conductivity coefficients $\sigma_{b/e}^{(i)},
i=1,2,3$, associated with the relevant tensors from (\ref{EE14}),
the dissipative rank-two tensors $\sigma_{b}^{\mu\nu}$ and
$\sigma_{e}^{\mu\nu}$ from (\ref{EE12}) read
\begin{eqnarray}\label{EE15}
\sigma_{i}^{\mu\nu}=
\sigma_{i}^{(1)}\Delta^{\mu\nu}+\sigma_{i}^{(2)}b^{\mu}b^{\nu}+
\sigma_{i}^{(3)}e^{\mu}e^{\nu},\qquad i=b,e.
\end{eqnarray}
Then, plugging $\sigma_{i}^{\mu\nu}, i=b,e$ from (\ref{EE15}) into
(\ref{EE12}), the dissipative part of the baryonic and
electric currents reads
\begin{eqnarray}\label{EE16}
j_{b}^{\mu}&=&-\sigma_{b}^{(1)}T\nabla^{\mu}\alpha_{b}-
\sigma_{b}^{(2)}b^{\mu}b^{\nu}T\nabla_{\nu}\alpha_{b}
-
\sigma_{b}^{(3)}e^{\mu}e^{\nu}T\nabla_{\nu}\alpha_{b}.\nonumber\\
j_{e}^{\mu}&=&-\sigma_{e}^{(1)}(T\nabla^{\mu}\alpha_{e}+E^{\mu})
-
\sigma_{e}^{(2)}b^{\mu}b^{\nu}(T\nabla_{\nu}\alpha_{e}+E_{\nu})-\sigma_{e}^{(3)}e^{\mu}e^{\nu}(T\nabla_{\nu}\alpha_{e}+E_{\nu}).
\end{eqnarray}
The
coefficients $\sigma_{b}^{(i)}$ and $\sigma_{e}^{(i)}$,
$i=1,2$ are previously introduced in \cite{rischke2010}. The remaining coefficients arise once the effects of the external electric field are not neglected.
\par
Concerning the viscous stress tensor $\tau^{\mu\nu}$ from (\ref{EE12}), we apply,
as in \cite{rischke2010}, the Onsager's principle
 $$\eta^{\mu\nu\rho\sigma}(E,B)=\eta^{\rho\sigma\mu\nu}(E,-B),$$
on the rank-four
tensor $\eta^{\mu\nu\rho\sigma}$ appearing in (\ref{EE12}).
All relevant tensors, compatible with this principle and
expressed in terms of $u^{\mu}, b^{\mu}$ and $e^{\mu}$ as well as $g^{\mu\nu}, b^{\mu\nu}$ and $e^{\mu\nu}$
are listed in Table \ref{table1} in four different series. All these bases fulfill the orthogonality condition, and are symmetric under $\rho\leftrightarrow \sigma$.
The bases appearing in series I and II are previously introduced in \cite{rischke2010}. The new bases appearing in series III include only $e^{\mu}$ and $e^{\mu\nu}$, and those in series IV include both electric and magnetic fields. Using these bases, and following the method presented in \cite{rischke2011} (see Appendix \ref{appAA} for more detail), the viscous stress tensor is then given by
\begin{widetext}
\begin{eqnarray}\label{EE17}
\tau^{\mu\nu}&=&2\eta_{0}\left(w^{\mu\nu}-\frac{1}{3}\Delta^{\mu\nu}\theta\right)+
\eta_{B}^{(1)}\left(\Delta^{\mu\nu}
-\frac{3}{2}\Xi_{B}^{\mu\nu}\right)\left(\theta-\frac{3}{2}\phi_{B}\right)
-2\eta_{B}^{(2)}\left(\Xi_{B}^{\mu\rho}b^{\nu}b^{\sigma}+\Xi_{B}^{\nu\rho}b^{\mu}b^{\sigma}
\right)w_{\rho\sigma}\nonumber\\
&&+2\eta_{B}^{(3)}\left(\Xi_{B}^{\mu\rho}b^{\nu\sigma}+\Xi_{B}^{\nu\rho}b^{\mu\sigma}\right)w_{\rho\sigma}
+2\eta_{B}^{(4)}\left(b^{\mu\rho}b^{\nu}b^{\sigma}
+b^{\nu\rho}b^{\mu}b^{\sigma}\right)w_{\rho\sigma}+\frac{3}{2}\zeta_{B}^{\perp}\Xi_{B}^{\mu\nu}\phi_{B}+3\zeta_{B}^{\|}b^{\mu}b^{\nu}\varphi_{B}\nonumber\\
&&+\eta_{E}^{(1)}\left(\Delta^{\mu\nu}
-\frac{3}{2}\Xi_{E}^{\mu\nu}\right)\left(\theta-\frac{3}{2}\phi_{E}\right)-2\eta_{E}^{(2)}\left(\Xi_{E}^{\mu\rho}e^{\nu}e^{\sigma}+
\Xi_{E}^{\nu\rho}e^{\mu}e^{\sigma}\right)w_{\rho\sigma}+\frac{3}{2}\zeta_{E}^{\perp}\Xi_{E}^{\mu\nu}\phi_{E}+3\zeta_{E}^{\|}e^{\mu}e^{\nu}\varphi_{E}\nonumber\\
&&+\eta_{EB}^{(1)}\left(\Delta^{\mu\nu}-\frac{3}{2}\Xi_{E}^{\mu\nu}\right)\left(\theta-\frac{3}{2}\phi_{B}\right)+
\eta_{EB}^{(2)}\left(\Delta^{\mu\nu}-\frac{3}{2}\Xi_{B}^{\mu\nu}\right)\left(\theta-\frac{3}{2}\phi_{E}\right)+2\eta_{EB}^{(3)}(\Xi_{E}^{\mu\rho}b^{\nu\sigma}+\Xi_{E}^{\nu\rho}b^{\mu\sigma})w_{\rho\sigma}\nonumber\\
&&
+2\eta_{EB}^{(4)}(b^{\mu\rho}e^{\nu}e^{\sigma}+b^{\nu\rho}e^{\mu}e^{\sigma})w_{\rho\sigma}
+2\eta_{EB}^{(5)}\left(\omega^{\mu}_{~\sigma}b^{[\nu,}e^{\sigma]}+
\omega^{\nu}_{~\rho}b^{[\mu,}e^{\rho]}\right)
\nonumber\\
&&-2\eta_{EB}^{(6)}[(\Xi_{B}^{\mu\rho}e^{\nu}e^{\sigma}+\Xi_{B}^{\nu\rho}e^{\mu}e^{\sigma})
+(\Xi_{E}^{\mu\rho}b^{\nu}b^{\sigma}+\Xi_{E}^{\nu\rho}b^{\mu}b^{\sigma})+2
(e^{\mu}b^{\rho}-b^{\mu}e^{\rho})(e^{\nu}b^{\sigma}-b^{\nu}e^{\sigma})]w_{\rho\sigma}+\frac{3}{2}\zeta_{EB}^{(1)\perp}\Xi_{B}^{\mu\nu}\phi_{E}\nonumber\\
&&+\frac{3}{2}\zeta_{EB}^{(2)\perp}\Xi_{E}^{\mu\nu}\phi_{B}
+3\zeta_{EB}^{(1)\|}(e^{\mu}e^{\nu}\varphi_{B}+b^{\mu}b^{\nu}\varphi_{E})
+6\zeta_{EB}^{(2)\|}b^{(\mu,}e^{\nu)}\varphi_{EB}+\zeta_{EB}^{(3)\|}\left(2\Delta^{\mu\nu}\varphi_{EB}-
\theta b^{(\mu,}e^{\nu)}\right).
\end{eqnarray}
\end{widetext}
\par\noindent
Here, $\phi_{B/E}\equiv \Xi^{\mu\nu}_{B/E}w_{\mu\nu}$,
$\varphi_{B}\equiv b^{\mu}b^{\nu}w_{\mu\nu}$,
$\varphi_{E}\equiv e^{\mu}e^{\nu}w_{\mu\nu}$ and $\varphi_{EB}\equiv b^{\mu}e^{\nu}w_{\mu\nu}$. Moreover $A^{(\mu,}B^{\nu)}\equiv A^{\mu}B^{\nu}+A^{\nu}B^{\mu}$. As expected, $\tau^{\mu\nu}$ is symmetric under $\mu\leftrightarrow\nu$ and satisfies $u_{\mu}\tau^{\mu\nu}=0$. According to our descriptions in \ref{appAA}, shear viscosities $\eta_{0}, \eta_{B}^{(i)}, i=1,\cdots,4, \eta_{E}^{(i)}, i=1,2$ and $\eta_{EB}^{(i)}, i=1,\cdots,6$ correspond to traceless rank-two tensors, while the tensors proportional to bulk viscosities
$\zeta_{i}^{\perp}, \zeta_{i}^{\|}, i=E,B$ as well as $\zeta_{EB}^{(i)\perp}, i=1,2$ and $\zeta_{EB}^{(i)\|}, i=1,2,3$ have non-vanishing traces. Shear viscosities $\eta_{0}, \eta_{B}^{(i)}, i=1,\cdots,4$ as well as bulk viscosities $\zeta_{B}^{\perp}$ and $\zeta_{B}^{\|}$ appear originally in \cite{lifshitz} as well as in \cite{rischke2010}. Here, we have completed the list of dissipative transport coefficients by considering the additional effect of an external electric field. Let us also notice that all bases including the combination $e_{\mu}b^{\mu\nu}$ are excluded once the condition (\ref{NE8}) is taken into account.
\par
Multiplying at this stage the expressions on the r.h.s. of (\ref{EE16}) and (\ref{EE17}) with $T\nabla_{\mu}\alpha_{b}, (T\nabla_{\mu}\alpha_{e}+E_{\mu})$ and $w_{\mu\nu}$, respectively [see (\ref{EE10})], and requiring $T\partial_{\mu}s^{\mu}\geq 0$, we arrive at
\begin{widetext}
\begin{eqnarray}\label{EE26}
\lefteqn{
T\partial_{\mu}s^{\mu}=2\eta_{0}\left(w^{\mu\nu}-\frac{1}{3}\Delta^{\mu\nu}\theta\right)^{2}
+\eta_{B}^{(1)}\left(\theta-\frac{3}{2}\phi_{B}\right)^{2}
+2\eta_{B}^{(2)}(b_{\mu}b^{\rho}w_{\rho\nu}-b_{\nu}b^{\rho}w_{\rho\mu})(b^{\mu}b_{\sigma}w^{\sigma\nu}-b^{\nu}b_{\sigma}w^{\sigma\mu})
}\nonumber\\
&&+\eta_{E}^{(1)}\left(\theta-\frac{3}{2}\phi_{E}\right)^{2}
+2\eta_{E}^{(2)}(e_{\mu}e^{\rho}w_{\rho\nu}-e_{\nu}e^{\rho}w_{\rho\mu})(e^{\mu}e_{\sigma}w^{\sigma\nu}-e^{\nu}e_{\sigma}w^{\sigma\mu})
\nonumber\\
&&
+2\eta_{EB}^{(6)}(b_{\mu}e^{\rho}w_{\rho\nu}-b_{\nu}e^{\rho}w_{\rho\mu})(b^{\mu}e_{\sigma}w^{\sigma\nu}-b^{\nu}e_{\sigma}w^{\sigma\mu})+2\eta_{EB}^{(6)}(e_{\mu}b^{\rho}w_{\rho\nu}-e_{\nu}b^{\rho}w_{\rho\mu})(e^{\mu}b_{\sigma}w^{\sigma\nu}-e^{\nu}b_{\sigma}w^{\sigma\mu})\nonumber\\
&&
+4\eta_{EB}^{(6)}\left(b^{\mu}e^{\rho}w_{\rho\sigma}-e^{\mu}b^{\rho}w_{\rho\sigma}\right)\left(b^{\sigma}e^{\nu}w_{\nu\mu}-e^{\sigma}b^{\nu}w_{\nu\mu}\right)
+\frac{3}{2}\zeta_{B}^{\perp}(\phi_{B}+\phi_{E})^{2}
+3\zeta_{B}^{\|}(\varphi_{B}+\varphi_{E})^{2}
+12\zeta_{EB}^{(2)\|}\varphi_{EB}^{2}
\nonumber\\
&&+(\sigma_{b}^{(1)}g^{\mu\nu}+\sigma_{b}^{(2)}b^{\mu}b^{\nu}
+\sigma_{b}^{(3)}e^{\mu}e^{\nu})
(T\nabla_{\mu}\alpha_{b})(T\nabla_{\nu}\alpha_{b})\nonumber\\
&&+(\sigma_{e}^{(1)}g^{\mu\nu}+\sigma_{e}^{(2)}b^{\mu}b^{\nu}
+\sigma_{e}^{(3)}e^{\mu}e^{\nu})
(T\nabla_{\mu}\alpha_{e}+E_{\mu})(T\nabla_{\nu}\alpha_{e}+E_{\nu}).
\end{eqnarray}
\end{widetext}
\par\noindent
In order to build a non-negative expression for $T\partial_{\mu}s^{\mu}$, we have set $\eta_{EB}^{(1)}=-\eta_{EB}^{(2)}$, and chosen $\zeta_{E}^{\perp}=\zeta_{EB}^{(1)\perp}=\zeta_{EB}^{(2)\perp}=\zeta_{B}^{\perp}$ as well as $\zeta_{E}^{\|}=\zeta_{EB}^{(1)\|}=\zeta_{B}^{\|}$. We also observe that $\eta_{B}^{(i)}, i=3,4$ as well as $\eta_{EB}^{(i)}, i=3,4,5$ and $\zeta_{EB}^{(3)\|}$ do not contribute to the entropy production.
\begin{center}
\begin{table*}[t]
\centering
\begin{tabular}{lcl}\hline
Series I&$\qquad$&Series II\\
\hline
$\Delta^{\mu\nu}\Delta^{\rho\sigma}$& $\qquad$&$\Delta^{\mu\nu}b^\rho b^\sigma+\Delta^{\rho\sigma}b^\mu b^\nu$\\
$ \Delta^{\mu\rho}\Delta^{\nu\sigma}+\Delta^{\mu\sigma}\Delta^{\nu\rho}$&$\qquad$&$b^\mu b^\nu b^\rho b^\sigma$\\
&$\qquad$&$\Delta^{\mu\rho}b^\nu
b^\sigma+\Delta^{\nu\rho}b^\mu b^\sigma+\Delta^{\mu\sigma}b^\nu
b^\rho+\Delta^{\nu\sigma}b^\mu b^\rho$\\
&$\qquad$&$\Delta^{\mu\rho}b^{\nu\sigma}+
\Delta^{\nu\rho}b^{\mu\sigma}+\Delta^{\mu\sigma}b^{\nu\rho}+\Delta^{\nu\sigma}b^{\mu\rho}
$ \\
&$\qquad$& $b^{\mu\rho}b^\nu b^\sigma+b^{\nu\rho}
b^\mu b^\sigma+b^{\mu\sigma}b^\nu b^\rho+b^{\nu\sigma}b^\mu
b^\rho$\\
\hline
Series III&$\qquad$&Series IV\\
\hline
$\Delta^{\mu\nu}e^\rho e^\sigma+\Delta^{\rho\sigma}e^\mu e^\nu$&$\qquad$&$ e^\mu e^\nu b^\rho b^\sigma+ b^\mu b^\nu e^\rho e^\sigma$\\
$e^\mu e^\nu e^\rho e^\sigma$&$\qquad$&$ b^{\rho\mu}e^\nu e^\sigma
+ b^{\rho\nu}e^\mu e^\sigma+b^{\sigma\mu}e^\nu e^\rho  +b^{\sigma\nu}e^\mu e^\rho
$\\
$\Delta^{\mu\rho}e^\nu
e^\sigma+
\Delta^{\nu\rho}e^\mu e^\sigma+\Delta^{\mu\sigma}e^\nu
e^\rho+\Delta^{\nu\sigma}e^\mu e^\rho$&$\qquad$&$
b^{(\mu,}e^{\nu)}b^{(\rho,}e^{\sigma)}$
\\
&$\qquad$&$\Delta^{\mu\rho}b^{[\nu,}e^{\sigma]}+\Delta^{\nu\sigma}b^{[\mu,}
e^{\rho]}+\Delta^{\mu\sigma}b^{[\nu,}
e^{\rho]}+\Delta^{\nu\rho}b^{[\mu,}
e^{\sigma]}$
\\
&$\qquad$&$\Delta^{\mu\nu}b^{(\rho,}e^{\sigma)}-\Delta^{\rho\sigma}b^{(\mu,}e^{\nu)}$
\\
\hline
\end{tabular}
\caption{Relevant combinations contributing to $\eta^{\mu\nu\rho\sigma}$. They are all compatible with the Onsager's principle $\eta^{\mu\nu\rho\sigma}(E,B)=\eta^{\rho\sigma\mu\nu}(E,-B)$ and the orthonormality condition $u_{\mu}\eta^{\mu\nu\rho\sigma}=0$. Moreover, they are all symmetric under $\rho\leftrightarrow\sigma$.}\label{table1}
\end{table*}
\end{center}
\subsection{Anomalous transport coefficients}\label{sec3B}
\par\noindent
Let us now consider the remaining terms
\begin{eqnarray*}
-j_{a}^{\mu}T\partial_{\mu}\alpha_{a}-j_{e}^{\mu}(T\partial_{\mu}\alpha_{e}-E_{\mu})+\mu_{a}C E\cdot B
+T\partial_{\mu}(D_{\omega}\omega^{\mu}+D_{B}B^{\mu}+D_{E}E^{\mu}),
\end{eqnarray*}
appearing in $T\partial_{\mu}s^{\mu}$ from (\ref{EE10}). In what follows, we will determine the anomalous coefficients $D_{i}, \kappa_{i}$ and $\xi_{i}, i=B,E,\omega$, introduced in $j_{e}^{\mu}$ and $j_{a}^{\mu}$ from (\ref{EE12}).
\par
To do this, let us insert
$j_{a}^{\mu}$ and the anomalous part of  $j_{e}^{\mu}$ from (\ref{EE12}) into the above expression and set the resulting expression equal to zero. We arrive first at
\begin{eqnarray}\label{EE27}
&&\hspace{0cm}-T\xi_{\omega}\omega\cdot \partial\alpha_{a}-T\xi_{B}B\cdot\partial\alpha_{a}
-T\xi_{E}E\cdot\partial\alpha_{a}\nonumber\\
&&\hspace{0cm}-T\kappa_{\omega}\omega\cdot\partial\alpha_{e}
-T\kappa_{E}E\cdot\partial\alpha_{e}
-T\kappa_{B}B\cdot\partial\alpha_{e}\nonumber\\
&&\hspace{0cm}-\kappa_{B}B\cdot E-\kappa_{\omega}E\cdot \omega
-\kappa_{E}
E\cdot E+C\alpha_{a}T B\cdot
E\nonumber\\
&&\hspace{0cm}+T\omega\cdot\partial D_{\omega}+TB\cdot\partial D_{B}+TE\cdot\partial D_{E} \nonumber\\
&&\hspace{0cm}+TD_{\omega}\partial_{\mu}\omega^{\mu}+TD_{B}
\partial_{\mu}B^{\mu}+TD_{E}\partial_{\mu}E^{\mu}=0.
\end{eqnarray}
At this stage, we have to insert the corresponding expressions to $\partial_{\mu}\omega^{\mu}, \partial_{\mu}B^{\mu}$, and $\partial_{\mu}E^{\mu}$ into (\ref{EE27}), and after reordering the resulting expression in terms of independent bases
\begin{eqnarray}\label{x1}
1, B^{\mu}, E^{\mu}, \omega^{\mu}, B\cdot\omega, E\cdot \omega, E\cdot E,B\cdot E,\end{eqnarray}
determine their coefficients.
In Appendix \ref{appA}, we have determined the exact values of $\partial_{\mu}B^{\mu}, \partial_{\mu}E^{\mu}$
and $\partial_{\mu}\omega^{\mu}$ [see (\ref{appA11a}), (\ref{appA12a}) and  (\ref{appA22a})]. Linearizing these expressions in terms of independent bases (\ref{x1}), and keeping only the terms in the second-order derivative expansion, we arrive at
\begin{eqnarray}\label{EE28}
\partial_{\mu}B^{\mu}&\approx&b_{1}(E\cdot\omega)-b_{2}B^{\mu}\partial_{\mu}p-b_{3}(B\cdot E), \nonumber\\
\partial_{\mu}E^{\mu}&\approx&e_{1}(B\cdot\omega)-e_{2}E^{\mu}\partial_{\mu}p-e_{3}(E\cdot E)+e_{4}n_{e}\nonumber\\
\partial_{\mu}\omega^{\mu}&\approx&
v_{1}(E\cdot \omega)+v_{2}\omega^{\mu}\partial_{\mu}p,
\end{eqnarray}
with
\begin{eqnarray}\label{EE29}
\begin{array}{rclcrcl}
b_{1}&=&2,&~&b_{2}&=&(\epsilon+p)^{-1},\\
b_{3}&=&n_{e}(1+\chi_{e})^{-1}b_{2},&~&\\
e_{1}&=&-2(1-\chi_{m})(1+\chi_{e})^{-1},&~&e_{2}&=&b_{2},\\
e_{3}&=&n_{e}(1+\chi_{e})^{-1}b_{2},&~&
e_{4}&=&(1+\chi_{e})^{-1},\\
v_{1}&=&-2n_{e}(1+\chi_{e})^{-1}b_{2},&~& v_{2}&=&-2b_{2}.
\end{array}
\end{eqnarray}
Here, we have used the fact that $B^{\mu}, E^{\mu}, \omega^{\mu}$ as well as $M$ and $P$ are ${\cal{O}}(\partial)$ and $u_{\mu}, \epsilon$ and $p$ are ${\cal{O}}(1)$ [see also \cite{son2009} for the same power counting]. In (\ref{EE28}), we have kept only terms in ${\cal{O}}(\partial^{2})$, and discard all the remaining terms. This defines our second-order derivative expansion.
Let us also note that the only effect of the medium, within this second-order derivative expansion, is reflected in the appearance of non-vanishing electric and magnetic susceptibilities, $\chi_{e}$ and $\chi_{m}$ in (\ref{EE21}). Thus, setting $\chi_{e}=\chi_{m}=0$ in (\ref{EE28}), the results for $\partial_{\mu}B^{\mu}$ and $\partial_{\mu}\omega^{\mu}$ presented in \cite{son2009} for $\partial_{\mu}B^{\mu}$
and $\partial_{\mu}\omega^{\mu}$ are reproduced. Plugging now (\ref{EE28}) into (\ref{EE27}), and simplifying the resulting expressions, we arrive first at
\begin{widetext}
\begin{eqnarray}\label{EE32}
0&=&+TB^{\mu}[\partial_{\mu}D_{B}-b_{2}D_{B}\partial_{\mu}p-\kappa_{B}\partial_{\mu}\alpha_{e}
-\xi_{B}\partial_{\mu}\alpha_{a}]+TE^{\mu}[\partial_{\mu}D_{E}-e_{2}D_{E}
\partial_{\mu}p-\kappa_{E}\partial_{\mu}\alpha_{e}-\xi_{E}\partial_{\mu}
\alpha_{a}]\nonumber\\
&&+T\omega^{\mu}[\partial_{\mu}D_{\omega}+v_{2}D_{\omega}\partial_{\mu}p-
\kappa_{\omega}\partial_{\mu}\alpha_{e}-\xi_{\omega}\partial_{\mu}\alpha_{a}]
+T(B\cdot E)\left(-b_{3}D_{B}-\frac{\kappa_{B}}{T}+C\alpha_{a}\right)
\nonumber\\
&&+
T(B\cdot\omega)e_{1}D_{E}
+T(E\cdot\omega)\left(v_{1}D_{\omega}+b_{1}D_{B}-\frac{\kappa_{\omega}}{T}\right)
-T(E\cdot E)\left(e_{3}D_{E}+\frac{\kappa_{E}}{T}\right)+Te_{4}n_{e}D_{E}.
\end{eqnarray}
\end{widetext}
\par\noindent
Using then the fact that the bases (\ref{x1}) are linear independent,
the expressions in front of them can be set independently equal to
zero. We arrive immediately at $D_{E}=\kappa_{E}=\xi_{E}=0$, as expected from the symmetry reasons. We conclude that the coefficients proportional to $E^{\mu}$ in $s^{\mu}, j_{e}^{\mu}$ and $j_{a}^{\mu}$ do not receive any contribution from anomaly. All the other anomalous transport coefficients satisfy the following algebro-differential equations:
\begin{eqnarray}\label{EE33}
&&\partial_{\mu}D_{B}-\frac{D_{B}\partial_{\mu}p}{(\epsilon+p)}-\kappa_{B}\partial_{\mu}\alpha_{e}-\xi_{B}
\partial_{\mu}\alpha_{a}=0,\nonumber\\
&&\partial_{\mu}D_{\omega}-\frac{2D_{\omega}\partial_{\mu}p}{(\epsilon+p)}-
\kappa_{\omega}\partial_{\mu}\alpha_{e}-\xi_{\omega}\partial_{\mu}\alpha_{a}=0,\nonumber\\
&&\frac{n_{e}}{(1+\chi_{e})}\frac{D_{B}}{(\epsilon+p)}+\frac{\kappa_{B}}{T}-C\alpha_{a}=0,\nonumber\\
&&\frac{2n_{e}}{(1+\chi_{e})}\frac{D_{\omega}}{(\epsilon+p)}-2D_{B}+\frac{\kappa_{\omega}}{T}=0.
\end{eqnarray}
Let us reiterate at this stage that to derive the above algebro-differential equations a number of constraints as $e_{\mu}b^{\mu\nu}=0$ from (\ref{NE8}) as well as $e_{\mu}\partial_{\mu}b^{\mu\nu}=0, \partial_{\mu}\chi_{e}=0$ and $\chi_{e}\neq 1$ are made. These kinds of constraints, especially those related to (\ref{NE8}), are used to derive the thermodynamical equations in Sec. \ref{sec2}. The latter are then used in Appendix \ref{appA} to derive general expressions for $\partial_{\mu}B^{\mu},\partial_{\mu}E^{\mu}$ and $\partial_{\mu}\omega^{\mu}$ in (\ref{appA11a}), (\ref{appA12a}) and (\ref{appA22a}), respectively. Approximating these relations in an appropriate way (see above) leads to $D_{E}=\xi_{E}=\kappa_{E}=0$ as well as to (\ref{EE33}), whose solutions yield anomalous transport coefficients $D_{k},\xi_{k},\kappa_{k}$ with $k=B,\omega$. This describes the role played by these constraints to determine these anomalous transport coefficients in this paper.
In a medium with vanishing $\chi_{e}$, the above equations (\ref{EE33}) reduce to the equations appearing in
\cite{sadofyev2010},
\begin{eqnarray}\label{EE34}
&&\partial_{\mu}D_{B}-\frac{D_{B}\partial_{\mu}p}{(\epsilon+p)}-
\kappa_{B}\partial_{\mu}\alpha_{e}-\xi_{B}\partial_{\mu}\alpha_{a}=0,\nonumber\\
&&\partial_{\mu}D_{\omega}-\frac{2D_{\omega}\partial_{\mu}p}{(\epsilon+p)}
-\kappa_{\omega}\partial_{\mu}\alpha_{e}-\xi_{\omega}\partial_{\mu}\alpha_{a}=0,\nonumber\\
&&\frac{n_{e}D_{B}}{(\epsilon+p)}+\frac{\kappa_{B}}{T}-C\alpha_{a}=0,\nonumber\\
&&\frac{2n_{e}D_{\omega}}{(\epsilon+p)}-2D_{B}+
\frac{\kappa_{\omega}}{T}=0.
\end{eqnarray}
A comparison between (\ref{EE33}) and (\ref{EE34}) shows that, within this second-order approximation, only the algebraic equation receives contribution from $\chi_{e}$, the electric susceptibility of the medium. The magnetic susceptibility, $\chi_{m}$, plays thus no role in modifying the anomalous transport coefficients in this approximation. In what follows, we first consider the algebro-differential equations (\ref{EE34}), and solve them to determine $D_{k}$, $\xi_{k}$, $\kappa_{k}$ for $k=B,\omega$ in terms of thermodynamical quantities $\epsilon, p,\alpha_{e}$ and $\alpha_{a}$. We then present the solutions of (\ref{EE33}) in an electrically polarized hot and dense medium in the presence of an external magnetic field.
\subsubsection{Anomalous transport coefficients in a medium with vanishing $\chi_{e}$ and $\chi_{m}$}\label{sec3B1}
\par\noindent  To solve (\ref{EE34}), we use the same method as in
\cite{son2009, sadofyev2010}. Introducing
$\alpha_{i}=\frac{\mu_{i}}{T}, i=a,e$ and replacing
$d\mu_{i}=Td\alpha_{i}+\alpha_{i}dT$ in the Gibbs-{\cal{D}}uhem equation
$dp=sdT+\sum_{i=\{e,a\}}n_{i}d\mu_{i}$, arising from
$\epsilon+p=Ts+\sum_{i=\{e,a\}}n_{i}\mu_{i}$ and
\begin{eqnarray*}
d\epsilon=Tds+\sum_{i=\{e,a\}}\mu_{i}dn_{i},
\end{eqnarray*}
we arrive at
\begin{eqnarray}\label{EE35}
\hspace{0cm}\partial_{\mu}p=\frac{\epsilon+p}{T}\partial_{\mu}T
+n_{e}T\partial_{\mu}\alpha_{e}+n_{a}T\partial_{\mu}\alpha_{a}.
\end{eqnarray}
We therefore have,
\begin{eqnarray}\label{EE36}
&&\hspace{0cm}\left(\frac{dp}{dT}\right)_{\alpha_{e},\alpha_{a}}=\sum\limits_{i=\{e,a\}}(s+n_{i}\alpha_{i})=\frac{\epsilon+p}{T},\nonumber\\
&&\hspace{0cm}\left(\frac{dp}{d\alpha_{e}}\right)_{\alpha_{a},
T}=n_{e}T,\qquad
\left(\frac{dp}{d\alpha_{a}}\right)_{\alpha_{e},
T}=n_{a}T.
\end{eqnarray}
Plugging at this stage,
\begin{eqnarray}\label{EE37}
\partial_{\mu}D_{k}=\left(\frac{\partial
D_{k}}{\partial T}\right)\partial_{\mu}T+\left(\frac{\partial
D_{k}}{\partial\alpha_{e}}\right)\partial_{\mu}\alpha_{e}+\left(\frac{\partial
D_{k}}{\partial \alpha_{a}}\right)\partial_{\mu}\alpha_{a},
\end{eqnarray}
for
$k=B,\omega$ and $\partial_{\mu}p$ from (\ref{EE35}) into the first two differential equations in (\ref{EE34}), we obtain
\begin{eqnarray}\label{EE38}
\frac{\partial D_{B}}{\partial T}&=&\frac{D_{B}}{T}, \nonumber\\
\frac{\partial D_{B}}{\partial\alpha_{e}}&=&\frac{n_{e}D_{B}T}{\epsilon+p}+\kappa_{B},\nonumber\\
\frac{\partial D_{B}}{\partial\alpha_{a}}&=&\frac{n_{a}D_{B}T}{\epsilon+p}+\xi_{B},
\end{eqnarray}
as well as
\begin{eqnarray}\label{EE39}
\frac{\partial D_{\omega}}{\partial T}&=&\frac{2D_{\omega}}{T}, \nonumber\\
\frac{\partial D_{\omega}}{\partial\alpha_{e}}&=&\frac{2n_{e}D_{\omega}T}{\epsilon+p}+\kappa_{\omega},\nonumber\\
\frac{\partial D_{\omega}}{\partial\alpha_{a}}&=&\frac{2n_{a}D_{\omega}T}{\epsilon+p}+\xi_{\omega}.
\end{eqnarray}
\par\noindent
Combining the last two algebraic relations in (\ref{EE34}) with the differential equations from  (\ref{EE38}) and (\ref{EE39}), we obtain
\begin{eqnarray}\label{EE40}
\hspace{0cm}D_{B}=T[C\alpha_{e}\alpha_{a}+\gamma_{B}(\alpha_{a})],~~~ D_{\omega}=T^{2}[C\alpha_{e}^{2}\alpha_{a}+2\alpha_{e}\gamma_{B}(\alpha_{a})+\gamma_{\omega}(\alpha_{a})],
\end{eqnarray}
where $\gamma_{k}(\alpha_{a}), k=B,\omega$ are constants of integration \cite{sadofyev2010}.
To arrive at (\ref{EE40}), the identities (\ref{EE36}) have been used.
Plugging $D_{B}$ and $D_{\omega}$ from (\ref{EE40}) into (\ref{EE38}) and (\ref{EE39}), we arrive at
\begin{eqnarray}\label{EE42}
\kappa_{B}&=&C\mu_{a}\left(1-\frac{n_{e}\mu_{e}}{\epsilon+p}\right)-\frac{n_{e}T^{2}\gamma_{B}}{\epsilon+p},\nonumber\\
\xi_{B}&=&C\mu_{e}\left(1-\frac{n_{a}\mu_{a}}{\epsilon+p}\right)+T\left(\frac{\partial\gamma_{B}}{\partial\alpha_{a}}-\frac{n_{a}T\gamma_{B}}{\epsilon+p}\right), \nonumber\\
\kappa_{\omega}&=&2C\mu_{e}\mu_{a}\left(1-\frac{n_{e}\mu_{e}}{\epsilon+p}\right)+2T^{2}\bigg[\left(1-\frac{2n_{e}\mu_{e}}{\epsilon+p}\right)\gamma_{B}-\frac{n_{e}T\gamma_{\omega}}{\epsilon+p}\bigg],\nonumber\\
\xi_{\omega}&=&C\mu_{e}^{2}\left(1-\frac{2n_{a}\mu_{a}}{\epsilon+p}\right)+2\mu_{e}T\left(\frac{\partial\gamma_{B}}{\partial\alpha_{a}}-\frac{2n_{a}T\gamma_{B}}{\epsilon+p}\right)+T^{2}\left(
\frac{\partial\gamma_{\omega}}{\partial\alpha_{a}}-\frac{2n_{a}T\gamma_{\omega}}{\epsilon+p}
\right),
\end{eqnarray}
which are consistent with the results presented in \cite{son2009}, provided the integration constants $\gamma_{i}, i=1,2$ are set to be zero (see below).
Here, $C=\frac{e^{2}}{4\pi^{2}}$ is the coefficient of the axial anomaly from (\ref{NE28}). Let us reiterate that
the first term in $\kappa_{B}$ is the same coefficient of chiral
magnetic effect, arising originally in \cite{kharzeev2007, kharzeev2008}. Moreover,  $\kappa_{\omega}$ is the coefficient for chiral vortical effect \cite{erdmenger2008, banerjee2008, torabian2009, son2009}, and $\xi_{B}$ as well as $\xi_{\omega}$, the coefficients of chiral vortical as well as chiral vortical separation effects, appeared first in \cite{son2004, metlitski2005} as well as in \cite{liao2015}.\footnote{In \cite{son2009}, no difference between $\mu_{e}$ and $\mu_{a}$ is made. Here, the transport coefficients corresponding to $B^{\mu}$ and $\omega^{\mu}$ are called $\xi$ and $\xi_{\omega}$. }
\par
Let us also note that the contributions from gravitational anomaly to anomalous transport coefficients are not considered in the present work. These kinds of corrections are computed in \cite{landsteiner2011-1, landsteiner2011-2} using an appropriate Kubo formalism (see also \cite{gao2012} for a kinetic theory approach). They are shown to appear as additional $T^{2}$ dependent terms in $\kappa_{k},\xi_{k}, k=B,\omega$. These terms can also be interpreted as contributions from the aforementioned integration constants $\gamma_{B}$ and $\gamma_{\omega}$ \cite{oz2010}. Their determination turns out to be strongly frame dependent \cite{abbasi2016, megias2017} (see also \cite{landsteiner2016} for a recent review). In what follows, we will determine $D_{k}, \xi_{k}, \kappa_{k}$ with $k=B,\omega$ for the case of non-vanishing $\chi_{e}$. Requiring that these results lead to the corresponding expressions for $\chi_{e}=0$, new constants of integrations will be brought in connection with $\gamma_{B}$ and $\gamma_{\omega}$.  The contributions from gravitational anomaly will not be considered in this framework, as in the case of $\chi_{e}=0$.
\subsubsection{Anomalous transport coefficients in a medium with non-vanishing $\chi_{e}$ and $\chi_{m}$}\label{sec3B2}
\par\noindent
In this section, we will use the method introduced in the previous
part to solve the differential equations (\ref{EE33}) for a
system with non-vanishing $\chi_{e}$. To do this, we first use the Gibbs-{\cal{D}}uhem
relation (\ref{NE22}), or equivalently
\begin{eqnarray}\label{EE43}
\partial_{\mu}p=\frac{(\epsilon+p)}{T}\partial_{\mu}T+n_{e}
T\partial_{\mu}\alpha_{e}+n_{a}T\partial_{\mu}\alpha_{a}+M\partial_{\mu}B
+E\partial_{\mu}P,
\end{eqnarray}
as well as
\begin{eqnarray}\label{EE44}
\partial_{\mu}D_{k}=
\left(\frac{\partial D_{k}}{\partial T}\right)\partial_{\mu}T+
\left(\frac{\partial
D_{k}}{\partial\alpha_{e}}\right)\partial_{\mu}\alpha_{e}
+\left(\frac{\partial
D_{k}}{\partial{\alpha_{a}}}\right)\partial_{\mu}\alpha_{a}+\left(\frac{\partial D_{k}}{\partial
B}\right)\partial_{\mu}B +\left(\frac{\partial D_{k}}{\partial{P}}\right)\partial_{\mu}P,
\end{eqnarray}
with $k=B,\omega$, and then rewrite the differential equations appearing in (\ref{EE33}). We arrive again at (\ref{EE38}) and (\ref{EE39})
as well as at $\frac{\partial D_{k}}{\partial B}=\frac{\partial D_{k}}{\partial P}={\cal{O}}(\partial^{3})$ for $k=B,\omega$.
To solve these sets of equations together with the two algebraic equations in (\ref{EE33}), let us start with the second equation in (\ref{EE38}). Replacing  $\kappa_{B}=C\alpha_{a}T-\frac{n_{e}}{(1+\chi_{e})}\frac{D_{B}T}{(\epsilon+p)}$ from the first algebraic equation in (\ref{EE33}), and defining a new variable $w\equiv \epsilon+p$, we obtain for constant $\alpha_{a}$ and $T$,
\begin{eqnarray}\label{EE47}
\frac{\partial D_{B}}{\partial \alpha_{e}}-\frac{\chi_{e}}{(1+\chi_{e})}\frac{\partial w}{\partial\alpha_{e}}\frac{D_{B}}{w}-C\alpha_{a}T=0.
\end{eqnarray}
To arrive at (\ref{EE47}), we have used  $\frac{\partial p}{\partial\alpha_{e}}=n_{e}T$ from (\ref{EE43}). For constant electric susceptibility, $\chi_{e}$, the solution of (\ref{EE47}) reads
\begin{eqnarray}\label{EE48}
D_{B}=w^{\frac{\chi_{e}}{1+\chi_{e}}}\left(A+C\alpha_{a}T\int d\alpha_{e} w^{-\frac{\chi_{e}}{1+\chi_{e}}}\right).
\end{eqnarray}
The integration constant  $A=A(\alpha_{a},T,\chi_{e})$ is fixed by requiring that
$D_{B}$ from (\ref{EE48}) satisfies the first equation of (\ref{EE38}). We arrive at
\begin{eqnarray}\label{EE49}
D_{B}=w^{\frac{\chi_{e}}{1+\chi_{e}}}\left(T^{\frac{1}{1+\chi_{e}}}\bar{\gamma}_{B}+C\alpha_{a}T \int d\alpha_{e} w^{-\frac{\chi_{e}}{1+\chi_{e}}}\right),
\end{eqnarray}
where $\bar{\gamma}_{B}(\alpha_{a},\chi_{e})$ is a constant of integration. Requiring further that $D_{B}$ from (\ref{EE49}) for $\chi_{e}=0$ is given by $D_{B}$ from (\ref{EE40}), we obtain $\bar{\gamma}_{B}(\alpha_{a},\chi_{e}=0)=\gamma_{B}(\alpha_{a})$, with $\gamma_{B}$ the integration constant arising in $D_{B}$ from (\ref{EE40}). To arrive at (\ref{EE49}), we have used
\begin{eqnarray}\label{EE50}
T\frac{\partial }{\partial T}w^{\pm\frac{n\chi_{e}}{1+\chi_{e}}}=\pm\frac{n\chi_{e}}{1+\chi_{e}}w^{\pm\frac{n\chi_{e}}{1+\chi_{e}}},
\end{eqnarray}
with $n=1$, which arises from
$\frac{\partial p}{\partial T}=\frac{w}{T}$ from (\ref{EE43}). Plugging $D_{B}$ from (\ref{EE49}) into the first algebraic equation of (\ref{EE33}), we arrive at
\begin{eqnarray}\label{EE51}
\kappa_{B}=CT\alpha_{a}\left(1-\frac{n_{e}T}{1+\chi_{e}}w^{-\frac{1}{1+\chi_{e}}}\int d\alpha_{e} w^{-\frac{\chi_{e}}{1+\chi_{e}}}\right)-\frac{n_{e}\bar{\gamma}_{B}}{(1+\chi_{e})}T^{\frac{2+\chi_{e}}{1+\chi_{e}}}w^{-\frac{1}{1+\chi_{e}}},
\end{eqnarray}
which reduces to $\kappa_{B}$ from (\ref{EE42}) for $\chi_{e}=0$, provided $\bar{\gamma}_{B}(\alpha_{a},\chi_{e}=0)=\gamma_{B}(\alpha_{a})$ from (\ref{EE40}). Differentiating $D_{B}$ from (\ref{EE49}) with respect to $\alpha_{a}$ leads, according to (\ref{EE38}), to
\begin{eqnarray}\label{EE52}
\xi_{B}&=&CT\left(1-\frac{1}{(1+\chi_{e})}\frac{n_{a}\alpha_{a}T}{w}\right)w^{\frac{\chi_{e}}{1+\chi_{e}}}\int d\alpha_{e} w^{-\frac{\chi_{e}}{1+\chi_{e}}}
-\frac{Cn_{a}\alpha_{a}T^{2}\chi_{e}}{1+\chi_{e}}w^{\frac{\chi_{e}}{1+\chi_{e}}}\int d\alpha_{e} w^{-\frac{(1+2\chi_{e})}{1+\chi_{e}}}\nonumber\\
&&+T^{\frac{1}{1+\chi_{e}}}w^{\frac{\chi_{e}}{1+\chi_{e}}}\left(\frac{\partial\bar{\gamma}_{B}}{\partial\alpha_{a}}-\frac{1}{1+\chi_{e}}\frac{n_{a}T\bar{\gamma}_{B}}{w}\right).
\end{eqnarray}
Here, $\frac{\partial p}{\partial\alpha_{a}}=n_{a}T$ from (\ref{EE43}) is used. For $\chi_{e}=0$, $\xi_{B}$ from (\ref{EE52}) leads, as expected, to $\xi_{B}$ from (\ref{EE42}), if
$\bar{\gamma}_{B}|_{\chi_{e}=0}=\gamma_{B}$ and $\frac{\partial\bar{\gamma}_{B}}{\partial\alpha_{a}}|_{\chi_{e}=0}=\frac{\partial\gamma_{B}}{\partial\alpha_{a}}$.
\par
Plugging at this stage $\kappa_{\omega}=2D_{B}T-\frac{2n_{e}TD_{\omega}}{w(1+\chi_{e})}$ from the second algebraic equation of (\ref{EE33}) into the r.h.s. of the second differential equation of (\ref{EE39}), and using $\frac{\partial p}{\partial\alpha_{e}}=n_{e}T$ from (\ref{EE43}), we arrive at the differential equation
\begin{eqnarray}\label{EE53}
\frac{\partial D_{\omega}}{\partial\alpha_{e}}-\frac{2\chi_{e}}{(1+\chi_{e})}\frac{\partial w}{\partial \alpha_{e}}\frac{D_{\omega}}{w}-2TD_{B}=0,
\end{eqnarray}
whose solution for constant $\chi_{e}$ reads
\begin{eqnarray}\label{EE54}
D_{\omega}=w^{\frac{2\chi_{e}}{1+\chi_{e}}}\left(T^{\frac{2}{1+\chi_{e}}}\bar{\gamma}_{\omega}+2T\int d\alpha_{e} D_{B}(\alpha_{e},\alpha_{a},T,\chi_{e})w^{-\frac{2\chi_{e}}{1+\chi_{e}}}\right),
\end{eqnarray}
with $D_{B}$ from (\ref{EE49}). As in the previous case, the integration constant is chosen so that $D_{\omega}$ from (\ref{EE54}) satisfies the first differential equation in (\ref{EE38}). Moreover, for $\chi_{e}=0$,  $D_{\omega}$ from (\ref{EE54}) is given by $D_{\omega}$ from (\ref{EE40}) provided $\bar{\gamma}_{\omega}|_{\chi_{e}=0}=\gamma_{\omega}$.
To determine $\kappa_{\omega}$, we use the second algebraic equation in (\ref{EE33}),  and arrive at
\begin{eqnarray}\label{EE55}
\kappa_{\omega}=2D_{B}T-\frac{2n_{e}T}{(1+\chi_{e})}\frac{D_{\omega}}{w},
\end{eqnarray}
with $D_{B}$ from (\ref{EE49}) and $D_{\omega}$ from (\ref{EE54}). Setting $\chi_{e}=0$, and using $D_{B}(\chi_{e}=0)$ and $D_{\omega}(\chi_{e}=0)$ from (\ref{EE40}), we immediately arrive at $\kappa_{\omega}$ from (\ref{EE42}).
Finally, differentiating $D_{\omega}$ from (\ref{EE54}) with respect to $\alpha_{a}$, and using the last equation of (\ref{EE38}), we arrive at
\begin{eqnarray}\label{EE56}
\xi_{\omega}&=&\frac{2n_{a}T^{2}(1-\chi_{e})}{1+\chi_{e}}w^{\frac{2\chi_{e}}{1+\chi_{e}}}\int d\alpha_{e}D_{B}w^{-\frac{(1+3\chi_{e})}{1+\chi_{e}}}-\frac{2n_{a}T}{(1+\chi_{e})}\frac{D_{\omega}}{w}+2Tw^{\frac{2\chi_{e}}{1+\chi_{e}}}\int d\alpha_{e}\xi_{B}w^{-\frac{2\chi_{e}}{1+\chi_{e}}}\nonumber\\
&&+w^{\frac{2\chi_{e}}{1+\chi_{e}}}T^{\frac{2}{1+\chi_{e}}}\frac{\partial\bar{\gamma}_{\omega}}{\partial\alpha_{a}},
\end{eqnarray}
where $D_{B}, \xi_{B}$ and $D_{\omega}$ are given in (\ref{EE49}), (\ref{EE52}) and (\ref{EE54}), respectively. Assuming $\frac{\partial\bar{\gamma}_{\omega}}{\partial\alpha_{a}}|_{\chi_{e}=0}=\frac{\partial\gamma_{\omega}}{\partial\alpha_{a}}$, $\xi_{\omega}$ from (\ref{EE56}) reduces to $\xi_{\omega}$ from (\ref{EE42}), as expected.\footnote{Let us emphasize that the integrations over $\alpha_{e}$ in the analytical expressions for the anomalous coefficients $D_{k},\xi_{k},\kappa_{k}$ wih $k=B,\omega$ (\ref{EE49})-(\ref{EE56}) are indefinite.}
\section{Concluding remarks}\label{sec4}
\par\noindent
The anomaly induced effects on magnetized chiral fluids have attracted much attention in recent years. They are all characterized by non-dissipative vector and axial vector currents, which are proportional to either the background magnetic field or the vorticity of the medium. The proportionality factors, whose values are dictated by axial anomaly, represent non-dissipative transport coefficients. They arise naturally within relativistic hydrodynamics, as shown originally by Son and Surowka in \cite{son2009}. In this paper, we have extended the method previously used in \cite{rischke2010, rischke2011} to the case of non-vanishing electric field. In addition, an anomalous current has been considered, which includes anomalous transport coefficients as in \cite{son2009, sadofyev2010}. This brings our derivation in connection with quantum anomalies. In this way,
the work of Son and Surowka is generalized to the case of an electromagnetized chiral fluid, which linearly responses to the external electromagnetic field through finite magnetization and electric polarization. We have shown that, within certain approximation, the anomalous transport coefficients are, in particular, given in terms of the electric susceptibility of the medium. Other ingredients are the energy density and  thermodynamic pressure of the medium as well as electric and axial charge densities.   They are all functions of the temperature $T$, finite electric and axial chemical potential, $\mu_{e}$ and $\mu_{a}$ of the fluid, as well as external electric and magnetic fields, $E$ and $B$ acting on the fluid. As a by product, we have also determined the complete set of dissipative transport coefficients arising in the dissipative part of the electric current as well as the viscous stress tensor. This completes the set of coefficients previously obtained in \cite{rischke2010, rischke2011}, where the dissipative coefficients arising from the external electric field were neglected.
\par
This work can be extended in many ways. One possibility is to assume
a certain thermodynamic potential for a chiral QCD-like effective model in the presence of parallel electric and magnetic fields. Using standard thermodynamical relations, it is then possible to explicitly determine the energy density, pressure and electric susceptibility of this model in terms of a given set of thermodynamical parameters $T,\mu_{e},\mu_{a}, E$ and $B$. The non-dissipative anomalous transport coefficients, which are presented in this works by certain integration over $\alpha_{e}=\frac{\mu_{e}}{T}$, can then be determined by numerically performing these integrals for a given set of $T,\mu_{e},\mu_{a}, E$ and $B$. It would be interesting to explore, for instance, the dependence of the anomalous transport coefficients on this set of parameters, especially when the chiral model exhibits a chiral phase transition. The behavior of the anomalous transport coefficients in the vicinity of chiral critical point might be interesting for the physics of quark matter under extreme conditions, and may have phenomenological consequences in HIC experiments. We will postpone these kinds of studies to our future works.
\section{Acknowledgments}\label{sec6}
\par\noindent
N.S. thanks D. H. Rischke and A. Sedrakian for useful
discussions, and S. Zeynizadeh for collaboration during the early stages
of this work. Both authors thank N. Abbasi for useful discussion on gravitational anomaly.

\begin{appendix}

\section{Determination of $\partial_{\mu}B^{\mu}$, $\partial_{\mu}E^{\mu}$ and
$\partial_{\mu}\omega^{\mu}$}\label{appA}
\subsection{Determination of $\partial_{\mu}B^{\mu}$ and  $\partial_{\mu}E^{\mu}$}
\par\noindent
\setcounter{equation}{0}
Let us consider the fluid energy-momentum tensor from (\ref{NE18}).
Using the definitions of $\epsilon', p_{\perp}$ and $p_{\|}$, it can
equivalently be given as
\begin{eqnarray}\label{appA1a}
T_{F(0)}^{\mu\nu}=\epsilon u^{\mu}u^{\nu}-p\Delta^{\mu\nu}+EP(u^{\mu}u^{\nu}-e^{\mu}e^{\nu})+MB\Xi_{B}^{\mu\nu},
\end{eqnarray}
where $\Xi_{B}^{\mu\nu}=\Delta^{\mu\nu}+b^{\mu}b^{\nu}$ and
$\Delta^{\mu\nu}=g^{\mu\nu}-u^{\mu}u^{\nu}$. To determine $\partial_{\mu}B^{\mu}$ and $\partial_{\mu}E^{\mu}$, let us consider the  combination $$\Delta_{\rho\nu}\partial_{\mu}T^{\mu\nu}_{(0)}
=\Delta_{\rho\nu}\partial_{\mu}T^{\mu\nu}_{F(0)}
+\Delta_{\rho\nu}\partial_{\mu}T^{\mu\nu}_{EM}=0.$$
Using the relations (\ref{NE11})-(\ref{NE17}) as well as the properties (\ref{NE8}) and $u_{\mu} e^{\mu}=u_{\mu}
b^{\mu}=0$, we arrive first at
\begin{eqnarray}\label{appA2a}
\hspace{-0.5cm}\Delta_{\rho\nu}\partial_{\mu}T^{\mu\nu}_{F(0)}=(\epsilon'+p_{\perp})
{\cal{D}}u_{\rho}-\nabla_{\rho}p_{\perp}
-\partial_{\mu}(EP e^{\mu}e_{\rho}-BM
b^{\mu}b_{\rho})+u_{\rho}(P{\cal{D}}E-M{\cal{D}}B)+u_{\rho}\theta(EP-BM),
\end{eqnarray}
and
\begin{eqnarray}\label{appA3a}
\Delta_{\rho\nu}\partial_{\mu}T^{\mu\nu}_{EM}=-n_{e}E_{\rho}+F_{\mu\rho}\partial_{\sigma}M^{\sigma\mu},
\end{eqnarray}
where $\nabla_{\mu}=\Delta_{\mu\nu}\partial^{\nu}$,
$\theta=\partial_{\mu}u^{\mu}$ and $E^{\mu}=F^{\mu\nu}u_{\nu}$.
Combining then these two relations, we get
\begin{eqnarray}\label{appA4a}
&&(\epsilon'+p_{\perp}){\cal{D}}u_{\rho}-\nabla_{\rho}p_{\perp}-\partial_{\mu}(EP
e^{\mu}e_{\rho}-BM
b^{\mu}b_{\rho})+u_{\rho}(P{\cal{D}}E-M{\cal{D}}B)+u_{\rho}\theta(EP-BM)-E_{\rho}n_{e}\nonumber\\
&&+F_{\mu\rho}\partial_{\sigma}M^{\sigma\mu}=0.
\end{eqnarray}
Multiplying at this stage (\ref{appA4a}) with $B^{\rho}$, and using the definition of $F^{\mu\rho}$ from (\ref{NE5}), we obtain
 \begin{eqnarray}\label{appA5a}
(\epsilon'+p_{\perp})(B\cdot
{\cal{D}}u)-B^{\rho}\partial_{\rho}p_{\perp}
-B^{\rho}\partial_{\mu}(EPe^{\mu}e_{\rho}-BMb^{\mu}b_{\rho})-n_{e}(E\cdot B)-(E\cdot B)u_{\rho}\partial_{\mu}M^{\mu\rho}=0.
\end{eqnarray}
Plugging $B\cdot {\cal{D}}u=-u\cdot {\cal{D}}B$ from (\ref{NE14}) into this relation,  and using the following two relations
\begin{eqnarray}\label{appA6a}
E\cdot {\cal{D}}u=-\frac{2(1-\chi_{m})}{(1+\chi_{e})}(B\cdot
\omega)+\frac{n_{e}-E^{\mu}\partial_{\mu}\chi_{e}}{(1+\chi_{e})}
-\partial_{\mu}E^{\mu},
\end{eqnarray}
and
\begin{eqnarray}\label{appA7a}
u_{\rho}\partial_{\mu}M^{\mu\rho}=\frac{2(\chi_{e}+\chi_{m})( B\cdot \omega)}{(1+\chi_{e})}-\frac{(E^{\mu}\partial_{\mu}\chi_{e}+n_{e}\chi_{e})}{(1+\chi_{e})},
\end{eqnarray}
we arrive after some straightforward computations at
\begin{widetext}
\begin{eqnarray}\label{appA8a}
\partial_{\mu}B^{\mu}+\frac{\chi_{e}(E\cdot
B)}{\epsilon'+p}\partial_{\mu}E^{\mu}&=&\frac{(\epsilon'+p_{\perp})}{(\epsilon'+p)}\bigg\{2(E\cdot
\omega)-\frac{B^{\rho}}{(\epsilon'+p_{\perp})}\left(\partial_{\rho}p_{\perp}+\frac{n_{e}E_{\rho}}{(1+\chi_{e})}+B\partial_{\rho}M
\right.\nonumber\\
&&\left.
+
E^{\mu}\partial_{\mu}(\chi_{e}E_{\rho})
+\frac{2E_{\rho}(\chi_{e}+\chi_{m})(B\cdot\omega)}{(1+\chi_{e})}-\frac{E_{\rho}E^{\mu}\partial_{\mu}\chi_{e}}{(1+\chi_{e})}\right)\bigg\}.
\end{eqnarray}
Multiplying at this stage (\ref{appA4a}) with $E^{\rho}$, we arrive first at
\begin{eqnarray}\label{appA9a}
(\epsilon'+p_{\perp})(E\cdot
{\cal{D}}u)-E^{\rho}\partial_{\rho}p_{\perp}-E^{\rho}\partial_{\mu}(EPe^{\mu}e_{\rho}-BMb^{\mu}b_{\rho})
-n_{e}(E\cdot E)-(E\cdot E)u_{\rho}\partial_{\mu}M^{\mu\rho}=0.
\end{eqnarray}
Then, using (\ref{appA6a}) and (\ref{appA7a}), we obtain
\begin{eqnarray}\label{appA10a}
\lefteqn{\hspace{0cm}\partial_{\mu}E^{\mu}-\frac{\chi_{m}(E\cdot
B)}{\epsilon+p_{\perp}}\partial_{\mu}B^{\mu}=-\frac{(\epsilon'+p_{\perp})}{(\epsilon+p_{\perp})}
\bigg\{\left(\frac{2(1-\chi_{m})}{(1+\chi_{e})}(B\cdot
\omega)-\frac{n_{e}}{(1+\chi_{e})}+\frac{E^{\mu}\partial_{\mu}\chi_{e}}{(1+\chi_{e})}\right)
}\nonumber\\
&&\hspace{0cm}+\frac{E^{\rho}}{\epsilon'+p_{\perp}}\left(\partial_{\rho}p_{\perp}+\frac{n_{e}E_{\rho}}{(1+\chi_{e})}-E\partial_{\rho}P-
B^{\mu}\partial_{\mu}(\chi_{m}B_{\rho})+\frac{2E_{\rho}(\chi_{e}+\chi_{m})(B\cdot\omega)}{(1+\chi_{e})}
-\frac{E_{\rho}E^{\mu}\partial_{\mu}\chi_{e}}{(1+\chi_{e})}\right)\bigg\}.
\end{eqnarray}
Combining now (\ref{appA8a}) with (\ref{appA10a}), we arrive at
\begin{eqnarray}\label{appA11a}
\lefteqn{
\partial_{\mu}B^{\mu}=\left(1+\frac{\chi_{e}\chi_{m}(E\cdot B)^{2}}{(\epsilon'+p)(\epsilon+p_{\perp})}\right)^{-1}\frac{(\epsilon'+p_{\perp})}{(\epsilon'+p)}
}\nonumber\\
&&\times\bigg\{2(E\cdot\omega)-\frac{B^{\rho}}{\epsilon'+p_{\perp}}\left(\partial_{\rho}p_{\perp}+E^{\mu}\partial_{\mu}(\chi_{e}E_{\rho})+B\partial_{\rho}M+\frac{n_{e}E_{\rho}}{(1+\chi_{e})}+\frac{2E_{\rho}(\chi_{e}+\chi_{m})(B\cdot \omega)}{(1+\chi_{e})}-\frac{E_{\rho}E^{\mu}\partial_{\mu}\chi_{e}}{(1+\chi_{e})}
\right)
\nonumber\\
&&
+\frac{\chi_{e}(E\cdot
B)}{(\epsilon+p_{\perp})}\bigg[\left(\frac{2(1-\chi_{m})}{(1+\chi_{e})}(B\cdot
\omega)-\frac{n_{e}}{(1+\chi_{e})}+\frac{E^{\rho}\partial_{\rho}\chi_{e}}{(1+\chi_{e})}\right)\nonumber\\
&&+\frac{E^{\rho}}{(\epsilon'+p_{\perp})}\left(\partial_{\rho}p_{\perp}+\frac{n_{e}E_{\rho}}{(1+\chi_{e})}-E\partial_{\rho}P-
B^{\mu}\partial_{\mu}(\chi_{m}B_{\rho})
+\frac{2E_{\rho}(\chi_{e}
+\chi_{m})(B\cdot \omega)}{(1+\chi_{e})}-\frac{E_{\rho}E^{\mu}\partial_{\mu}\chi_{e}}{(1+\chi_{e})}
\right)\bigg]\bigg\},
\end{eqnarray}
and
\begin{eqnarray}\label{appA12a}
\lefteqn{
\partial_{\mu}E^{\mu}=-\left(1+\frac{\chi_{e}\chi_{m}(E\cdot
B)^{2}}{(\epsilon'+p)(\epsilon+p_{\perp})}\right)^{-1}\frac{(\epsilon'+p_{\perp})}{(\epsilon+p_{\perp})}
\bigg\{
\frac{2(1-\chi_{m})}{(1+\chi_{e})}(B\cdot
\omega)-\frac{n_{e}}{(1+\chi_{e})}+\frac{E^{\rho}\partial_{\rho}\chi_{e}}{(1+\chi_{e})}
}\nonumber\\
&&+\frac{E^{\rho}}{(\epsilon'+p_{\perp})}\left(\partial_{\rho}p_{\perp}+\frac{n_{e}E_{\rho}}{(1+\chi_{e})}-E\partial_{\rho}P-
B^{\mu}\partial_{\mu}(\chi_{m}B_{\rho})+\frac{2E_{\rho}(\chi_{e}+\chi_{m})(B\cdot \omega)}{(1+\chi_{e})}-\frac{E_{\rho}E^{\mu}\partial_{\mu}\chi_{e}}{(1+\chi_{e})}
\right)\nonumber\\
&&-\frac{\chi_{m}(E\cdot
B)}{(\epsilon'+p)}\bigg[2(E\cdot\omega)-\frac{B^{\rho}}{(\epsilon'+p_{\perp})}\left(\partial_{\rho}p_{\perp}
+E^{\mu}\partial_{\mu}(\chi_{e}E_{\rho})+B\partial_{\rho}M+\frac{n_{e}E_{\rho}}
{(1+\chi_{e})}\right.\nonumber\\
&&\left.+\frac{2E_{\rho}(\chi_{e}+\chi_{m})(B\cdot\omega)}{(1+\chi_{e})}-\frac{E_{\rho}E^{\mu}\partial_{\mu}\chi_{e}}{(1+\chi_{e}}\right)
\bigg]\bigg\}.
\end{eqnarray}
\end{widetext}
\par\noindent
To prove (\ref{appA6a}), we have used the definition of $E^{\mu}=F^{\mu\nu}u_{\nu}$ to get
$E^{\rho}{\cal{D}}u_{\rho}=(u^{\mu}F^{\rho\nu})u_{\nu}\partial_{\mu}u_{\rho}$.
To determine $u^{\mu}F^{\rho\nu}$, we multiply
$B^{\sigma}=\frac{1}{2}\varepsilon^{\sigma\rho\nu\mu}F_{\rho\nu}u_{\mu}$
with $\varepsilon_{\sigma\rho'\nu'\mu'}$, and use
\begin{eqnarray}\label{appA13a}
\varepsilon^{\sigma}_{~\rho\nu\mu}\varepsilon_{\sigma\rho'\nu'\mu'}=-\{g_{\rho'\rho}(g_{\nu'\nu}g_{\mu'\mu}-g_{\nu'\mu}g_{\mu'\nu})
-g_{\rho'\nu}(g_{\nu'\rho}g_{\mu'\mu}-g_{\nu'\mu}g_{\mu'\rho})
+ g_{\rho'\mu}(g_{\nu'\rho}g_{\mu'\nu}-g_{\nu'\nu}g_{\mu'\rho})\},
\end{eqnarray}
to obtain
$F^{\rho\nu}u^{\mu}=-\varepsilon^{\sigma\rho\nu\mu}B_{\sigma}+F^{\rho\mu}u^{\nu}
+F^{\mu\nu}u^{\rho}$.
We arrive first at
\begin{eqnarray}\label{appA14a}
E\cdot
{\cal{D}}u=(u^{\mu}F^{\rho\nu})u_{\nu}\partial_{\mu}u_{\rho}=-2(B\cdot\omega)
+F^{\rho\mu}\partial_{\mu}u_{\rho}.
\end{eqnarray}
Here, $u_{\mu} u^{\mu}=1, u_{\rho}\partial_{\mu}u^{\rho}=0$ and the
definition of the vorticity
$\omega^{\sigma}=\frac{1}{2}\varepsilon^{\sigma\rho\nu\mu}u_{\nu}\partial_{\mu}u_{\rho}$ are used.
Considering $F^{\rho\mu}\partial_{\mu}u_{\rho}$ on the r.h.s. of (\ref{appA14a}), and replacing $F^{\mu\rho}$ with $F^{\mu\rho}=H^{\mu\rho}+M^{\mu\rho}$, we arrive, upon using the equation of motion (\ref{NE9}), at
$\partial_{\mu}F^{\mu\rho}=n_{e(0)}^{\rho}+\partial_{\mu}M^{\mu\rho}$. Plugging this relation into (\ref{appA14a}), we obtain
\begin{eqnarray}\label{appA15a}
\hspace{-0.5cm}E\cdot {\cal{D}}u=-2(B\cdot\omega)-\partial_{\mu}E^{\mu}+n_{e}+u_{\mu}\partial_{\sigma}
M^{\sigma\mu}.
\end{eqnarray}
Using the definition of $M^{\mu\rho}$ from  (\ref{NE5}), and performing some straightforward computations, we obtain
\begin{eqnarray}\label{appA16a}
u_{\rho}\partial_{\mu}M^{\mu\rho}=-E^{\mu}\partial_{\mu}\chi_{e}-\chi_{e}\partial_{\mu}E^{\mu}+\chi_{e}u\cdot {\cal{D}}E +2\chi_{m}(B\cdot\omega).
\end{eqnarray}
Here, the definition of magnetic susceptibility
$\chi_{m}=\frac{M}{B}$ is used. Combining (\ref{appA15a}) with (\ref{appA16a}), we arrive finally at (\ref{appA6a}) and (\ref{appA7a}).
\subsection{Determination of $\partial_{\mu}\omega^{\mu}$}
\par\noindent
To determine $\partial_{\mu}\omega^{\mu}$, let us multiply
(\ref{appA4a}) with $\omega^{\rho}$ to arrive first at
\begin{eqnarray}\label{appA17a}
(\epsilon'+p_{\perp})(\omega\cdot
{\cal{D}}u)-\omega^{\rho}\partial_{\rho}p_{\perp}-n_{e}(E\cdot\omega)
+\omega^{\rho}\partial_{\mu}
(BMb^{\mu}b_{\rho}-EPe^{\mu}e_{\rho})+\omega^{\rho}F_{\mu\rho}\partial_{\sigma}M^{\sigma\mu}=0.
\end{eqnarray}
To determine $\omega^{\rho}{\cal{D}}u_{\rho}$, we use the definition of the vorticity
$\omega^{\rho}=\frac{1}{2}\varepsilon^{\rho\mu\alpha\beta}u_{\mu}\partial_{\alpha}u_{\beta}$,
and rewrite the combination $\eta_{\mu\lambda\rho}\equiv
u_{\mu}\partial_{\lambda}u_{\rho}$ in
$\omega^{\rho}{\cal{D}}u_{\rho}=\frac{1}{2}\varepsilon^{\rho\mu\alpha\beta}u^{\lambda}\partial_{\alpha}
u_{\beta}\eta_{\mu\lambda\rho}$ in terms of the vorticity $\omega^{\mu}$.
We obtain
\begin{eqnarray}\label{appA18a}
\eta_{\mu\lambda\rho}=-2\varepsilon_{\xi\mu\lambda\rho}\omega^{\xi}+\eta_{\mu\rho\lambda}+\eta_{\lambda\mu\rho}-
\eta_{\lambda\rho\mu}-\eta_{\rho\mu\lambda}+\eta_{\rho\lambda\mu}.
\end{eqnarray}
Plugging $\eta_{\mu\lambda\rho}$ from (\ref{appA18a}) into
\begin{eqnarray*}
\omega^{\rho}{\cal{D}}u_{\rho}=\frac{1}{2}\varepsilon^{\rho\mu\alpha\beta}u^{\lambda}\partial_{\alpha}
u_{\beta}\eta_{\mu\lambda\rho},
\end{eqnarray*}
and using symmetry arguments as well as the normalization property
of $u_{\mu}$, we obtain
\begin{eqnarray}\label{appA19a}
u\cdot D\omega=\frac{1}{2}\partial_{\mu}\omega^{\mu}.
\end{eqnarray}
To determine $\omega^{\rho}F_{\mu\rho}\partial_{\sigma}M^{\sigma\mu}$, we use the definitions of $\omega^{\rho}$ and $F_{\mu\rho}$, and after a lengthy but straightforward computation, where in particular
\begin{eqnarray}\label{appA20a}
B^{\sigma\mu}B_{\mu\rho}&=&-B^{2}\Xi_{B~\rho}^{\sigma},\nonumber\\
\omega^{\rho}B_{\mu\rho}&=&\frac{1}{2}\big[-u_{\mu}B\cdot {\cal{D}}u+B^{\rho}\partial_{\mu}u_{\rho}-B^{\rho}\partial_{\rho}u_{\mu}\big],\nonumber\\
B_{\mu\rho}\partial_{\sigma}B^{\sigma\mu}&=&B^{\sigma}\partial_{\rho}B_{\sigma}-B^{\sigma}\partial_{\sigma}B_{\rho}+B_{\rho}B^{\sigma} {\cal{D}}u_{\sigma}+u_{\rho}B^{\sigma}u^{\alpha}\partial_{\sigma}B_{\alpha}-u_{\rho}B^{\sigma}{\cal{D}}B_{\sigma}-(B\cdot B){\cal{D}}u_{\rho},
\end{eqnarray}
are used, we arrive first at
\begin{eqnarray}\label{appA21a}
\omega^{\rho}F_{\mu\rho}\partial_{\sigma}M^{\sigma\mu}&=&-(E\cdot\omega)
u_{\mu}\partial_{\sigma}M^{\sigma\mu}
+(B\cdot B)\omega^{\rho}\partial_{\rho}\chi_{m}-(B\cdot\omega)B^{\rho}\partial_{\rho}\chi_{m}
\nonumber\\
&&-\frac{1}{2}\big[\chi_{e}(B\cdot {\cal{D}}u)(E\cdot {\cal{D}}u)-\chi_{e}E^{\sigma}B^{\rho}(\partial_{\mu}u_{\rho}-\partial_{\rho}u_{\mu})\partial_{\sigma}u^{\mu}\nonumber\\
&&+\chi_{e}B^{\rho}{\cal{D}}E^{\mu}(\partial_{\mu}u_{\rho}-\partial_{\rho}u_{\mu})\big]+\chi_{m}B^{\sigma}\omega^{\rho}
\partial_{\rho}B_{\sigma}\nonumber\\
&&-\chi_{m}\omega^{\rho}B^{\sigma}\partial_{\sigma}B_{\rho}+\chi_{m}(B\cdot\omega)(B\cdot {\cal{D}}u)-\chi_{m}(B\cdot B)(\omega\cdot {\cal{D}}u).
\end{eqnarray}
Adding this expression to the remaining terms in (\ref{appA17a}), and using (\ref{NE14}), (\ref{appA6a}) and (\ref{appA7a}) as well as (\ref{appA19a}), we arrive finally at
\begin{widetext}
\begin{eqnarray}\label{appA22a}
\partial_{\mu}\omega^{\mu}&=&-\frac{2}{\epsilon'+p}\bigg\{\omega^{\rho}\big[\partial_{\rho}p_{\perp}+\chi_{e}E^{\mu}
\partial_{\mu}
E_{\rho}
-\chi_{m}B^{\sigma}\partial_{\rho}B_{\sigma}-(B\cdot B)~\partial_{\rho}\chi_{m}\big]+n_{e}(E\cdot \omega)\nonumber\\
&&+\frac{\chi_{e}}{2}\partial_{\mu}B^{\mu}\big[\partial_{\rho}E^{\rho}-\frac{(n_{e}-E^{\rho}\partial_{\rho}\chi_{e})}{1+\chi_{e}}\big]
+\frac{\chi_{e}(1-\chi_{m})}{1+\chi_{e}}\partial_{\mu}B^{\mu}(B\cdot\omega)\nonumber\\
&&-\frac{\chi_{e}}{2}(B^{\rho}E^{\sigma}\partial_{\sigma}u^{\mu}-B^{\rho}{\cal{D}}E^{\mu})(\partial_{\mu}u_{\rho}-\partial_{\rho}u_{\mu})\bigg\}.
\end{eqnarray}
\end{widetext}
\par\noindent
Here, $\partial_{\mu}B^{\mu}$ and $\partial_{\mu}E^{\mu}$ are to be read from (\ref{appA11a}) and (\ref{appA12a}), respectively. In Section \ref{sec3B}, we will use a certain approximation in the second-order of derivative expansion to simplify these relations as well as $\partial_{\mu}\omega^{\mu}$ from (\ref{appA22a}). The resulting expressions (\ref{EE28}) will then be used to determine the anomalous transport coefficients $D_{k},\xi_{k}$ as well as $\kappa_{k}$ with $k=B,\omega,E$.

\section{Relevant bases for the shear and bulk viscosities from (\ref{EE17})}\label{appAA}
\par\noindent
In this section we present the bases used to build the rank-two tensors appearing in $\tau^{\mu\nu}$ from (\ref{EE17}).
\par
In order to build the coefficients of the shear viscosities $\eta_{0}$ and $\eta_{B}^{(i)}, i=1,2,3$, following traceless combinations are used \cite{rischke2010}
\begin{eqnarray}\label{EE18}
&a)&\Delta^{\mu\rho}\Delta^{\nu\sigma}+\Delta^{\mu\sigma}\Delta^{\nu\rho}
-\frac{2}{3}\Delta^{\mu\nu}\Delta^{\rho\sigma},\nonumber\\
&b)&\left(\Delta^{\mu\nu}-\frac{3}{2}\Xi_{B}^{\mu\nu}\right)\left(\Delta^{\rho\sigma}-\frac{3}{2}\Xi_{B}^{\rho\sigma}\right),\nonumber\\
&c)&-\left(\Xi_{B}^{\mu\rho}b^{\nu}b^{\sigma}+\Xi_{B}^{\nu\rho}b^{\mu}b^{\sigma}
+\Xi_{B}^{\mu\sigma}b^{\nu}b^{\rho}+\Xi_{B}^{\nu\sigma}b^{\mu}b^{\rho}\right),
\nonumber\\
&d)&\Xi_{B}^{\mu\rho}b^{\nu\sigma}+\Xi_{B}^{\nu\rho}b^{\mu\sigma}
+\Xi_{B}^{\mu\sigma}b^{\nu\rho}+\Xi_{B}^{\nu\sigma}b^{\mu\rho},
\nonumber\\
&e)&b^{\mu\rho}b^{\nu}b^{\sigma}+b^{\nu\rho}b^{\mu}b^{\sigma}
+b^{\mu\sigma}b^{\nu}b^{\rho}+b^{\nu\sigma}b^{\mu}b^{\rho}.
\end{eqnarray}
In addition to these combinations, there are other relevant traceless combinations, leading to the coefficients for $\eta_{E}^{(i)}, i=1,2$
\begin{eqnarray}\label{EE19}
&a)& \left(\Delta^{\mu\nu}-\frac{3}{2}\Xi_{E}^{\mu\nu}\right)
\left(\Delta^{\rho\sigma}-\frac{3}{2}\Xi_{E}^{\rho\sigma}\right),\nonumber\\
&b)&-\left(\Xi_{E}^{\mu\rho}e^{\nu}e^{\sigma}+\Xi_{E}^{\nu\rho}e^{\mu}e^{\sigma}
+\Xi_{E}^{\mu\sigma}e^{\nu}e^{\rho}+\Xi_{E}^{\nu\sigma}e^{\mu}e^{\rho}\right),
\end{eqnarray}
and to $\eta_{EB}^{(i)}, i=1,\cdots,6$
\begin{eqnarray}\label{EE20}
&a)& \left(\Delta^{\mu\nu}-\frac{3}{2}\Xi_{E}^{\mu\nu}\right)
\left(\Delta^{\rho\sigma}-\frac{3}{2}\Xi_{B}^{\rho\sigma}\right),\nonumber\\
&b)& \left(\Delta^{\mu\nu}-\frac{3}{2}\Xi_{B}^{\mu\nu}\right)
\left(\Delta^{\rho\sigma}-\frac{3}{2}\Xi_{E}^{\rho\sigma}\right),\nonumber\\
&c)&\Xi_{E}^{\mu\rho}b^{\nu\sigma}+\Xi_{E}^{\nu\rho}b^{\mu\sigma}
+\Xi_{E}^{\mu\sigma}b^{\nu\rho}+\Xi_{E}^{\nu\sigma}b^{\mu\rho},\nonumber\\
&d)&b^{\mu\rho}e^{\nu}e^{\sigma}+b^{\nu\rho}e^{\mu}e^{\sigma}
+b^{\mu\sigma}e^{\nu}e^{\rho}+b^{\nu\sigma}e^{\mu}e^{\rho},\nonumber\\
&e)& \Delta^{\mu\rho}b^{[\nu,}e^{\sigma]}+\Delta^{\nu\sigma}b^{[\mu,}
e^{\rho]}+\Delta^{\mu\sigma}b^{[\nu,}
e^{\rho]}+\Delta^{\nu\rho}b^{[\mu,}
e^{\sigma]},\nonumber\\
&f)&-\Xi_{B}^{\mu\rho}e^{\nu}e^{\sigma}-\Xi_{B}^{\nu\rho}e^{\mu}e^{\sigma}
-\Xi_{B}^{\mu\sigma}e^{\nu}e^{\rho}-\Xi_{B}^{\nu\sigma}e^{\mu}e^{\rho}-\Xi_{E}^{\mu\rho}b^{\nu}b^{\sigma}-\Xi_{E}^{\nu\rho}b^{\mu}b^{\sigma}
-\Xi_{E}^{\mu\sigma}b^{\nu}b^{\rho}-\Xi_{E}^{\nu\sigma}b^{\mu}b^{\rho}
\nonumber\\
&&-4 (e^{\mu}e^{\nu}b^{\rho}b^{\sigma}+b^{\mu}b^{\nu}e^{\rho}e^{\sigma})+2(b^{\mu}e^{\nu}+b^{\nu}e^{\mu})(b^{\rho}e^{\sigma}+b^{\sigma}e^{\rho}).
\end{eqnarray}
Let us notice that the combinations $c)$ and $d)$ from (\ref{EE20}) are only traceless when the assumption $e_{\mu}b^{\mu\nu}=0$ from (\ref{NE8}) is taken into account. Otherwise, only the sum of these two combinations will be traceless.
\par
To build the bulk viscosities $\zeta_{B/E}^{\perp}$ and $\zeta_{B/E}^{\|}$ following combinations with non-vanishing trace are used
\begin{eqnarray}\label{EE21}
\begin{array}{lrclr}
a)&\Xi_{B}^{\mu\nu}\Xi_{B}^{\rho\sigma},&\qquad&
b)&\Xi_{E}^{\mu\nu}\Xi_{E}^{\rho\sigma},\nonumber\\
c)&b^{\mu}b^{\nu}b^{\rho}b^{\sigma},&\qquad&
d)&e^{\mu}e^{\nu}e^{\rho}e^{\sigma}.
\end{array}
\end{eqnarray}
Other relevant combinations with non-vanishing trace are
\begin{eqnarray}\label{EE23}
a)~\Xi_{B}^{\mu\nu}\Xi_{E}^{\rho\sigma},\qquad
b)~\Xi_{E}^{\mu\nu}\Xi_{B}^{\rho\sigma},
\end{eqnarray}
and
\begin{eqnarray}\label{EE24}
&a)&e^{\mu}e^{\nu}b^{\rho}b^{\sigma}+b^{\mu}b^{\nu}e^{\rho}e^{\sigma}\nonumber\\
&b)&(b^{\mu}e^{\nu}+b^{\nu}e^{\mu}),(b^{\rho}e^{\sigma}+b^{\sigma}e^{\rho}),\nonumber\\
&c)&\Delta^{\mu\nu}b^{(\rho,}e^{\sigma)}-\Delta^{\rho\sigma}b^{(\mu,}e^{\nu)}.
\end{eqnarray}
These combinations are used to build the coefficients $\zeta_{EB}^{(i)\perp}, i=1,2$ and $\zeta_{EB}^{(i)\|}, i=1,2,3$ appearing in (\ref{EE19}).

\end{appendix}

\end{document}